\documentclass[final,5p,times,twocolumn]{elsarticle}

\usepackage{amssymb}
\usepackage{amsmath}

\usepackage{amsthm}
\usepackage{framed} 
\usepackage{multicol} 
\usepackage{tikz}
\usetikzlibrary{arrows.meta, positioning}
\usepackage{xcolor}
\usepackage{graphicx}

\usepackage{graphics}
\usepackage{enumitem}
\newtheorem{theorem}{Theorem}
\newtheorem{remark}{Remark} 
\usepackage{subcaption}

\usepackage{siunitx}

\usepackage{booktabs}

\usepackage{nomencl} 
\makenomenclature
\renewcommand*\nompreamble{\begin{multicols}{2}}
	\renewcommand*\nompostamble{\end{multicols}}
	
\usepackage[hidelinks]{hyperref}

\begin{document}
	
	\makeatletter
	\def\ps@pprintTitle{}
	\makeatother

\begin{frontmatter}

 \title{Fault-Tolerant Temperature Control of HRSG Superheaters: Stability Analysis Under Valve Leakage Using Physics-Informed Neural Networks}

\author[inst1]{Mojtaba Fanoodi}

\author[inst1]{Farzaneh Abdollahi}

\author[inst2]{Mahdi Aliyari Shoorehdeli}

\author[inst3]{Mohsen Maboodi}

\affiliation[inst1]{
	organization={Department of Electrical Engineering, AmirKabir University of Technology (Tehran Polythechnique)},
	city={Tehran},
	country={Iran}
}

\affiliation[inst2]{
	organization={Department of Electrical Engineering, K.N. Toosi University of Technology},
	city={Tehran},
	country={Iran}
}

\affiliation[inst3]{
	organization={Digitalization SBU Management, Mapna Electric \& Control Engineering \& Manufacturing (MECO)},
	city={Alborz},
	country={Iran}
}

\begin{abstract}
Faults and operational disturbances in Heat Recovery Steam Generators (HRSGs), such as valve leakage, present significant challenges, disrupting steam temperature regulation and potentially causing efficiency losses, safety risks, and unit shutdowns. Traditional PI controllers often struggle due to inherent system delays, nonlinear dynamics, and static gain limitations. This paper introduces a fault-tolerant temperature control framework by integrating a PI plus feedforward control strategy with Physics-Informed Neural Networks (PINNs). The feedforward component anticipates disturbances, preemptively adjusting control actions, while the PINN adaptively tunes control gains in real-time, embedding thermodynamic constraints to manage varying operating conditions and valve leakage faults. A Lyapunov-based stability analysis confirms the asymptotic convergence of temperature tracking errors under bounded leakage conditions. Simulation results using operational data from the Pareh-Sar combined cycle power plant demonstrate significantly improved response times, reduced temperature deviations, enhanced fault resilience, and smooth gain adjustments. The proposed adaptive, data-driven methodology shows strong potential for industrial deployment, ensuring reliable operation, autonomous fault recovery, and enhanced performance in HRSG systems.

\end{abstract}


\begin{keyword}

Heat Recovery Steam Generator (HRSG) \sep Steam Temperature Control \sep Feedforward Control Strategy \sep Physics-Informed Neural Network (PINN)  \sep Leakage Fault

\end{keyword}

\end{frontmatter}

\section{Introduction}
\label{sec1}

A Heat Recovery Steam Generator (HRSG) is a key component of combined-cycle power plants, recovering waste heat from gas-turbine exhaust to generate high-pressure steam for additional power production \cite{arpit_state---art_2023}. This process substantially increases overall plant efficiency, often exceeding 60\%. Thermodynamic modeling plays a vital role in optimizing HRSG design and predicting system performance under varying operating conditions \cite{zhang_thermodynamic_2016, erdogan_integrated_2025}.

Among HRSG components, the superheater and desuperheater are central to achieving stable and efficient steam temperature regulation \cite{hennessey_6_2017}. The superheater raises steam temperature to improve turbine efficiency, while the desuperheater prevents overheating through precise water injection \cite{wang_new_2013}. Accurate modeling of these components enables reliable prediction of transient behavior and identification of optimal control strategies \cite{zima_simulation_2019}. Recent studies highlight the importance of evaluating their coupled dynamics for enhanced safety and energy efficiency in real operations.

One significant challenge in HRSG modeling is managing disturbances like variations in gas turbine exhaust temperatures due to load changes, directly affecting steam temperature control. Traditional feedback control methods react only after errors occur, causing delays and instability. Recent approaches employ advanced control strategies, such as fractional-order PID and fuzzy controllers, improving dynamic responses \cite{elhosseini_heat_2022}. Waste heat recovery technologies reviewed across multiple industries emphasize the importance of reliable control systems to fully realize the benefits of energy recovery and minimize process variability \cite{jouhara_waste_2018}.

\begin{figure}[h!]
	\centering
	\begin{subfigure}[t]{0.45\columnwidth}
		\centering
		\includegraphics[width=\linewidth]{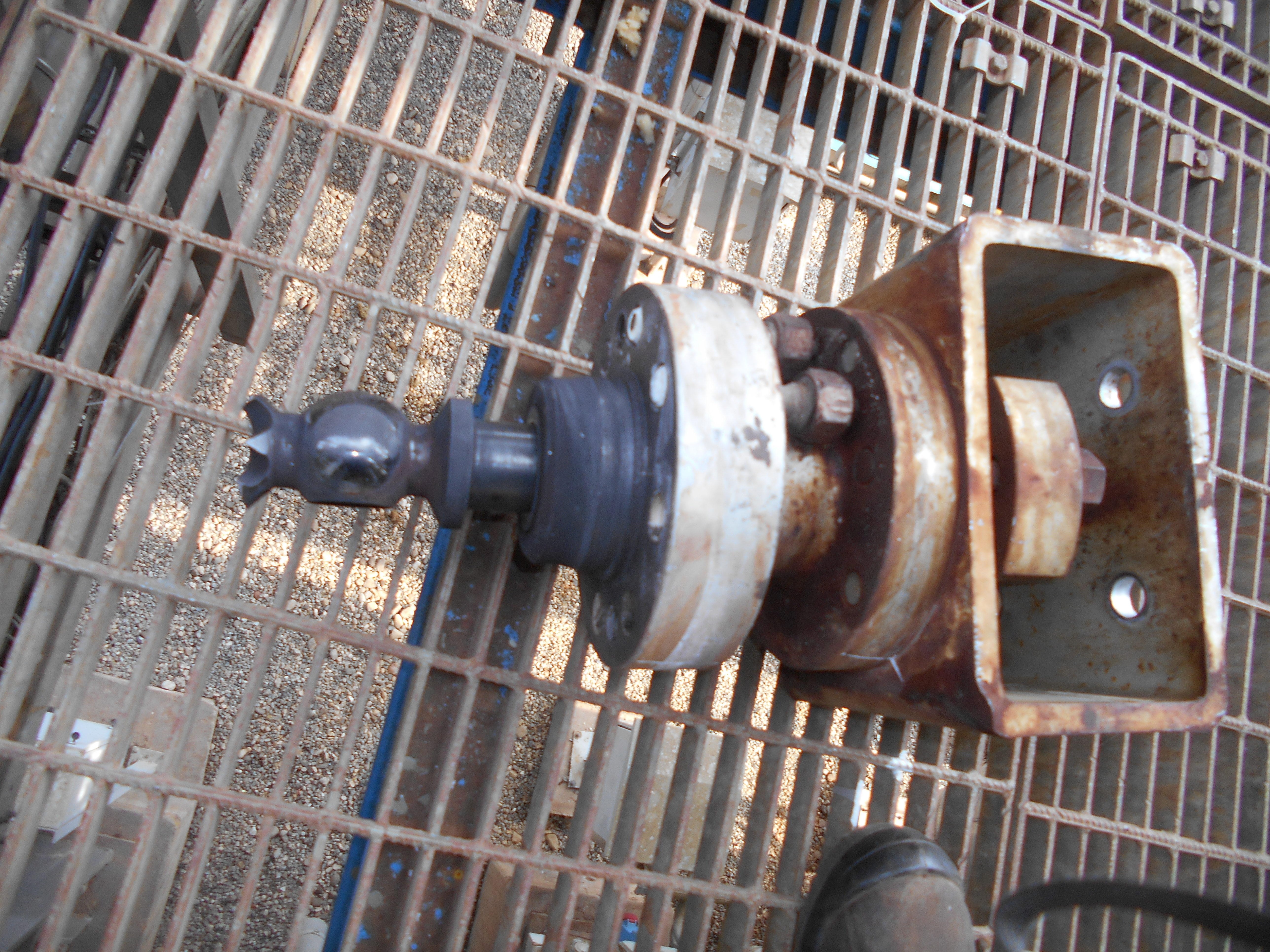}
		\caption{}
	\end{subfigure}%
	\hfill
	\begin{subfigure}[t]{0.45\columnwidth}
		\centering
		\includegraphics[width=\linewidth]{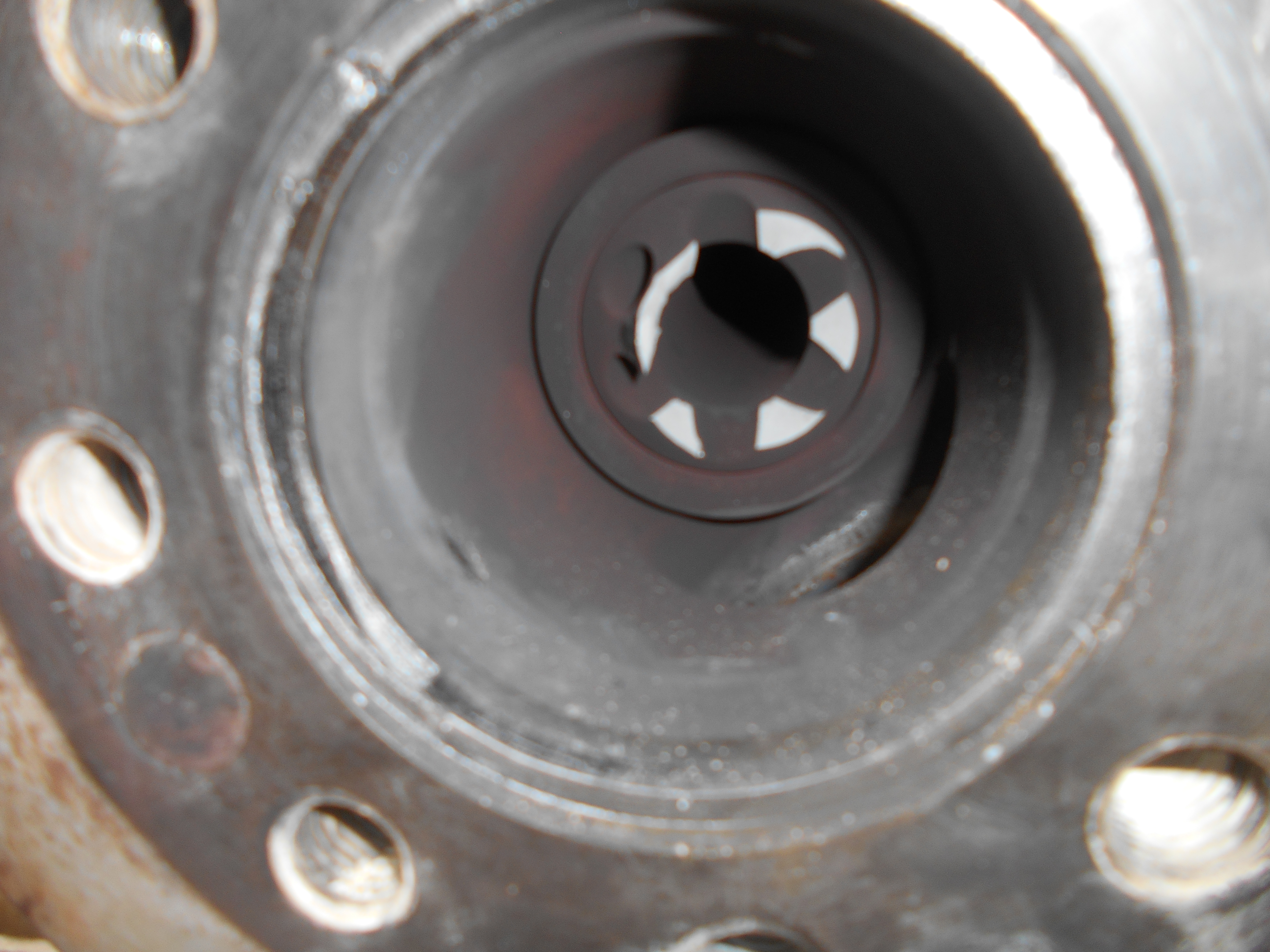}
		\caption{}
	\end{subfigure}
	\caption{Valve leakage in nozzles of superheaters and its effect in Pareh-Sar Power Plant, located in northern Iran.
		a: DSH nozzle taken down for maintenance. b: Cracks inside the nozzle.}
	\label{fig01}
\end{figure}

\begin{figure}[h!]
	\centering
	\includegraphics[width=0.65\columnwidth]{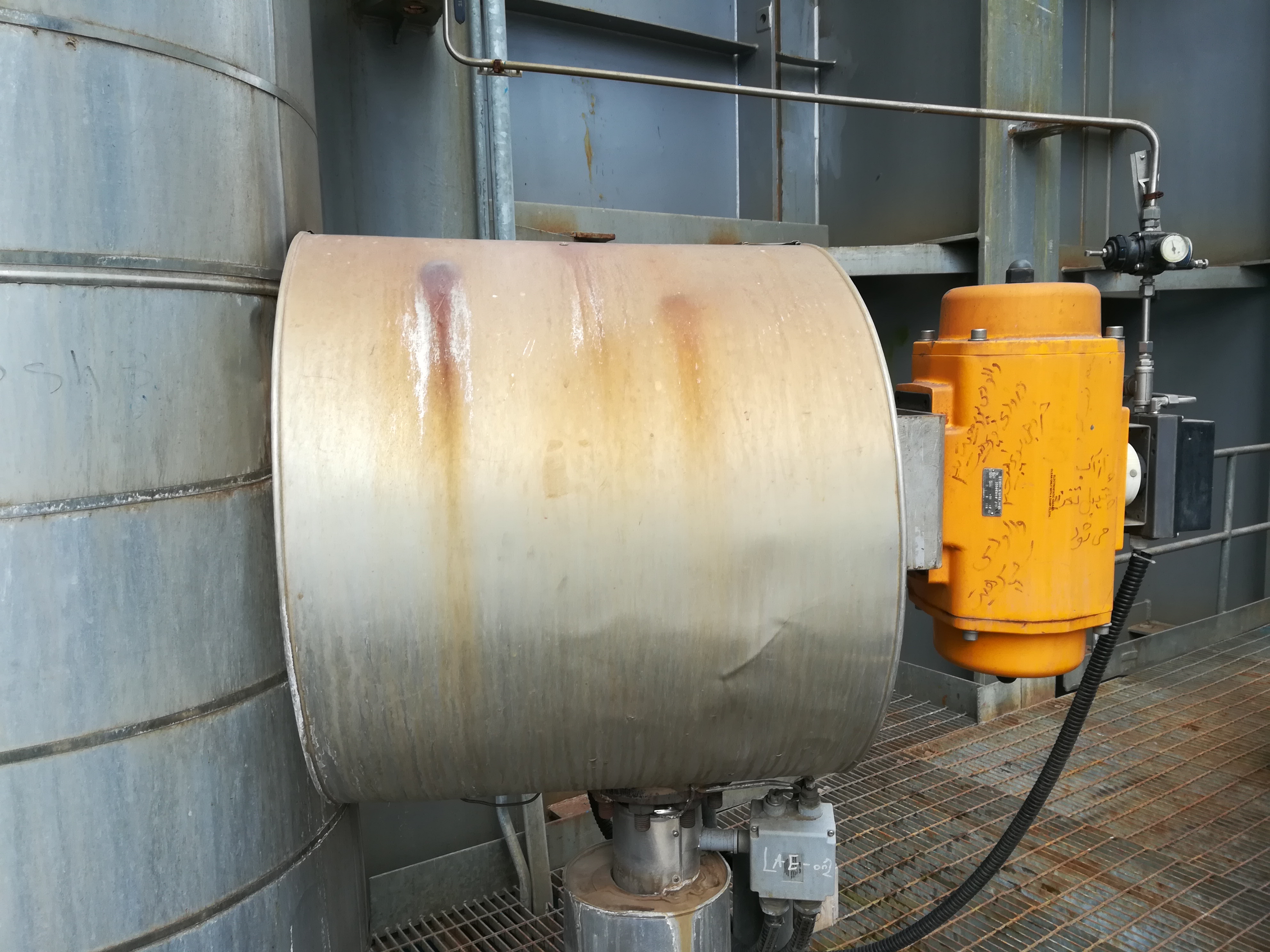}
	\caption{DSH of the HRSG in Pareh-Sar Powerplant }
	\label{fig02}
\end{figure}

Valve leakage faults, particularly in the desuperheater, complicate HRSG temperature regulation. These faults, caused by mechanical wear or actuator malfunctions, significantly disrupt temperature control, reduce efficiency, and increase safety risks \cite{zhang_failure_2020}. Even minor leakage can substantially impact temperature regulation, necessitating emergency shutdowns and costly manual interventions. For instance, an undetected valve leakage at Pareh-Sar Power Plant in 2022 led to severe temperature overshoot, triggering emergency shutdowns and incurring significant expenses. Examples of valve leakage and associated damage are illustrated in Figures \ref{fig01} and \ref{fig02}.

\nomenclature{$C$}{Specific heat [$\frac{kJ}{kgK}$]}
\nomenclature{$h$}{Specific enthalpy [$\frac{kJ}{kg}$]}
\nomenclature{$\bar{h}$}{Enthalpy size [$\frac{kJ}{mol}$]}
\nomenclature{$T$}{Temperature [Celsius degrees]}
\nomenclature{$t$}{Time [Seconds]}
\nomenclature{$\bar{m}$}{Mass [$\frac{kJ}{kgK}$]}
\nomenclature{$\rho$}{Density [$\frac{kg}{m^3}$]}
\nomenclature{$m$}{Weight [kg]}
\nomenclature{$\dot{m}$}{Mass flow rate [$\frac{kg}{m}$]}
\nomenclature{$Q$}{Heat exchanged [MJ]}
\nomenclature{$H$}{Caloric value [$\frac{kJ}{kg}$]}
\nomenclature{$V$}{Volume [$m^3$]}
\nomenclature{\mbox{}}{}

\begin{table*}[!t]
	\centering
	\begin{framed}
		\small
		\printnomenclature
	\end{framed}
	\vspace{-2mm}
\end{table*}

Moreover, superheater and desuperheater components are vulnerable to erosion and corrosion, exacerbating valve leakage issues and system inefficiencies \cite{kochmanski_failure_2024}\cite{li_failure_2022}. Regular monitoring and advanced diagnostic tools are critical for early detection and mitigation of such faults. For instance, the use of bi-sensor information fusion has been demonstrated to improve leakage detection accuracy, while early detection of anomalies through machine learning techniques enables proactive measures to prevent cascading failures \cite{liu_comprehensive_2024}\cite{Yang_2021}\cite{zhao_anomaly_2018}.

In a similar work, the effectiveness of analytical redundancy techniques, particularly the residue method, has been demonstrated for fault detection in steam superheater systems \cite{maican_application_2023}. Similarly, adaptive observer-based fault-tolerant control methods have been successfully applied in thermofluid systems, where combined backstepping control and adaptive fault estimation maintained stable temperature regulation under both dynamic and sensor faults \cite{han2023dynamic}. These studies highlight the potential of adaptive control strategies combined with predictive diagnostics to enhance the robust performance of HRSG. Furthermore, valve leakage can lead to operational inefficiencies, increased emissions, and even safety hazards due to internal flow disturbances \cite{jafari_valve_2014}\cite{qin_internal_2023}. Additionally, Flow Accelerated Corrosion (FAC) has been identified as a significant contributor to internal leakage within HRSGs. Turbulence and water quality can exacerbate FAC, leading to leakage issues in control valves and also in economizers. Regular maintenance and stringent monitoring of water quality are thus critical in minimizing FAC and its adverse effects on system performance \cite{kumar_2019}.

Advanced adaptive fault-tolerant control techniques have also been developed for nonlinear systems affected by actuator faults and dead-zones, where neural network–based approximation and prescribed-performance backstepping ensure finite-time stability under switching behaviors \cite{bali2023adaptive}. Recent studies have demonstrated the effectiveness of adaptive and gain-scheduling control strategies in improving HRSG temperature regulation under nonlinear and time-varying conditions. These methods dynamically adjust controller gains based on operating points or estimated parameters, outperforming fixed-gain designs in transient response and steady-state accuracy \cite{nakamoto1995multivariable, kang_2013, begum_design_2016}. More recent approaches integrate neural and physics-informed models for adaptive gain tuning, offering improved robustness and interpretability \cite{bolderman2024physics}.

In contrast, conventional fixed-gain PI controllers remain limited in handling faults such as valve leakage \cite{saha_automatic_2017}. Historical data from the Pareh-Sar power plant show temperature overshoots of up to 6–8~\si{\degreeCelsius} during leakage events, occasionally triggering safety alarms and manual intervention. Similar challenges appear in gas-turbine control, where override logic and Min–Max schemes require rigorous stability analysis due to their hybrid nature, highlighting the importance of ensuring absolute stability under switching behaviors \cite{imani2020stability}. While model-predictive and machine-learning-based controllers have been explored to enhance performance, their complexity and computational cost restrict real-time deployment in industrial HRSGs.

 These limitations motivate the development of a physics-informed adaptive control framework capable of providing real-time gain adjustment with provable stability and low computational overhead.
To address these issues, this paper proposes a novel fault-tolerant control framework integrating a PI plus feedforward strategy with Physics-Informed Neural Networks (PINNs). PINNs dynamically adjust control gains $(K_p, K_i, K_{ff})$ in real-time, embedding thermodynamic laws and enhancing robustness. Additionally, a Lyapunov-based stability analysis confirms asymptotic convergence of temperature tracking errors under bounded valve leakage conditions.

The rest of the paper is structured as follows. Section 2 details the superheater and desuperheater model development, providing comprehensive modeling of the HRSG system components. Section 3 introduces the control strategy, incorporating the PI plus feedforward approach. In section 4, we elaborate on the Development of a PINN along with the associated Lyapunov stability analysis for adaptive gain tuning. Section 5 presents simulation and real time site results, validating the proposed control methodology against operational data from the Pareh-Sar power plant, highlighting enhanced system response and fault resilience. Finally, section 6 concludes the paper by summarizing key findings and suggesting future research directions.

\section{Superheater and Desuperheater Model Development}
\subsection{Superheater}
Past studies, like \cite{chaibakhsh_modelling_2013}, emphasizes the importance of accurately simulating HRSG performance over extended operational periods to enhance the understanding of their dynamics and efficiency. They employed a dynamic modeling framework that incorporates various operational parameters and thermodynamic principles to simulate the behavior of HRSGs under different conditions. This model allows for the analysis of heat transfer processes, pressure drops, and temperature variations within the HRSG, providing insights into how these factors influence overall system performance. Their model also incorporated linearization and imposed restrictive conditions on thermodynamic relations, while using a genetic algorithm to estimate parameters. Despite its limitations, it describes the thermodynamic equation of the outlet temperature of superheater section as follows\cite{chaibakhsh_modelling_2013}:
\begin{equation}
	\small
	\frac{d}{dt}\!\left[m_a C_p T_a\right] 
	+ \rho_s V_s C_p \frac{dT_{\mathrm{out_{SH}}}}{dt} 
	= Q + C_p \dot{m}_{\mathrm{in_{SH}}} \left(T_{\mathrm{in_{SH}}} - T_{\mathrm{out_{SH}}}\right)
\end{equation}

where: \( m_a \) is the mass of metallic parts (kg), \( C_p \) is the specific heat capacity of air (kJ/kgK), \( T_a \) the temperature of metallic parts (\si{\degreeCelsius}), \( \rho_s \) is the density of the material in the superheater (kg/$m^3$), \( V_s \) is the volume of the superheater ($m^3$), \( T_{out_{SH}} \) is the outlet temperature of the superheater (\si{\degreeCelsius}), \( Q \) is the heat exchanged (MJ), \( \dot{m}_{in_{SH}} \) is the mass flow rate at the inlet (kg/s), \( T_{in_{SH}} \) is the inlet temperature (\si{\degreeCelsius}).

The heat transferred can be expressed in terms of the fuel's calorific value and as a function of the fuel flow rate as:

\begin{equation}
	Q = H \cdot \dot{m}_{\text{fuel}}
\end{equation}

where \( H \) represents the calorific value(kJ/kg) and \( \dot{m}_{\text{fuel}} \) is the mass flow rate of fuel (kg/s). Thus, by substituting and rewriting the aforementioned equation, we have:

\begin{equation}
	\dot{T}_{\text{out}_{\text{SH}}} = K_2 \left( K_1 \dot{m}_{fuel} + \dot{m}_{in_{SH}} (T_{in_{SH}} - T_{out_{SH}}) - \dot{\widetilde{T}}_a \right)
\end{equation}

where:
\[
K_1 = \frac{H}{C_p} \quad \quad \quad K_2 = \frac{1}{\rho_s V_s} \quad \quad \quad \dot{\widetilde{T}}_a = \frac{d}{dt} \left[ m_a \times C_p \times T_a \right]
\]

where \( \dot{T}_{\text{out}_{\text{SH}}} \) is the rate of change of outlet temperature in the superheater (\si{\degreeCelsius}/s), \( \dot{\widetilde{T}}_a \) is the rate of change in the temperature of metallic parts, and \( \dot{m}_{\text{out}_{\text{DSH}}} \) is the mass flow rate at the outlet of the desuperheater (kg/s).

On the other hand, considering the law of energy conservation and the system structure, we know that:

\begin{equation}
	\dot{m}_{\text{in}\ \text{SH}} = \dot{m}_{\text{out}\ \text{DSH}}
\end{equation}

Finally, the differential equation for the superheater output temperature is rewritten as:

\begin{equation}
	\dot{T}_{\text{out}_{\text{SH}}} = K_2 \left( K_1 \dot{m}_{fuel} + \dot{m}_{\text{out}\ \text{DSH}} (T_{in_{SH}} - T_{out_{SH}}) - \dot{\widetilde{T}}_a \right)
\end{equation}

\subsection{De-superheater}

The desuperheater is used to reduce the temperature of the superheated steam by injecting cooling water or spray water. The temperature control of the desuperheater is crucial for maintaining safe and efficient plant operation. The dynamics are more complex due to the mixing of steam and spray water, which can be captured using mass and energy balance principles.
The mass flow rate of steam at the outlet of the desuperheater is the sum of the inlet steam mass flow rate and the mass flow rate of the spray water:
\begin{equation}
	\dot{m}_{out_{dsh}} = \dot{m}_{in_{dsh}} + \dot{m}_{spray}
\end{equation}

where:
\(\dot{m}_{out_{dsh}}\) is the outlet mass flow rate of steam after the desuperheater,
\(\dot{m}_{in_{dsh}}\) is the inlet mass flow rate of superheated steam,
\(\dot{m}_{spray}\) is the mass flow rate of the injected spray water.

The change in steam temperature where it passes through the desuperheater is calculated using the conservation of energy, taking into account the temperature of both the incoming steam and the spray water\cite{maffezzoni_boiler-turbine_1997}:
\begin{equation}
	\begin{split}
		\dot{T}_{\mathrm{out_{dsh}}} 
		&= \frac{\left(T_{\mathrm{in_{dsh}}} - T_{\mathrm{out_{dsh}}}\right)
			\dot{m}_{\mathrm{out_{dsh}}}}{\bar{m}_{\mathrm{out_{dsh}}}} \quad + \frac{\bar{m}_{\mathrm{in_{dsh}}}}{\bar{m}_{\mathrm{out_{dsh}}}}
		\dot{T}_{\mathrm{in_{dsh}}} \\
		&\quad - \frac{T_{\mathrm{in_{dsh}}} - T_{\mathrm{spray}}}
		{\bar{m}_{\mathrm{out_{dsh}}}}
		\dot{m}_{\mathrm{spray}}
	\end{split}
\end{equation}

where:
\(T_{out_{dsh}}\) is the outlet steam temperature after the desuperheater,
\(T_{in_{dsh}}\) is the inlet temperature of the superheated steam,
\(T_{spray}\) is the temperature of the injected spray water,
\(\bar{m}_{out}\) is the average mass flow rate at the outlet.

This equation shows how the outlet temperature of the desuperheater is influenced by the temperature difference between the inlet steam and the spray water, as well as the mass flow rates of the steam and spray water.

A key relationship in the desuperheater model is between temperature and enthalpy. The change in enthalpy (\(\Delta h\)) can be related to the change in temperature (\(\Delta T\)) using the specific heat capacity at constant pressure:

\begin{equation}
	\Delta h = C_p \Delta T
\end{equation}

This relationship is used to compute the temperature variations based on the energy input or output within each component of the HRSG. For instance, during the desuperheating process, as the spray water is mixed with the superheated steam, the enthalpy changes reduce in the steam temperature.

Building on the mass and energy balance equations, we can refine the expression for the output temperature of the desuperheater by incorporating both mass flow dynamics and energy conservation:

\begin{equation}
	\dot{T}_{out_{dsh}} = \frac{T_{in_{dsh}} - T_{out_{dsh}}}{\bar{m}_{out_{dsh}}}\dot{m}_{out} + \frac{\bar{m}_{in_{dsh}}}{\bar{m}_{out_{dsh}}}\dot{T}_{in_{dsh}} - \frac{T_{in} - T_{spray}}{\bar{m}_{out_{dsh}}}\dot{m}_{spray}
\end{equation}

This formula provides precise calculation of the outlet temperature under varying conditions, considering the mass flow rate of steam and spray water and the energy exchange during the process.
The primary objective of the control strategy is to maintain the steam temperature at desired levels. In the HRSG system, controlling the temperature at the outlet of the superheater and desuperheater is critical for ensuring the efficient and safe operation of the power plant. To achieve this goal, we develop a model that relates the manipulated variables (spray water injection) to the controlled output (steam temperature).
The primary control action is achieved through by adjusting the spray water flow rate into the desuperheater.
To design a control strategy, we represent the system in state-space representation. The state-space model captures the relationship between the input (spray water injection) and the output (steam temperature) through first-order differential equations.

Let the state variables be defined as:
\(x_1(t)\): The steam temperature at superheater outlet.
\(x_2(t)\): The steam temperature at desuperheater outlet.

The input to the system is the spray water flow rate \(u(t)\), and the disturbances include variations in gas turbine temperature and inlet steam conditions. The state-space representation is given by:

\begin{equation}
	\begin{cases}
		\dot{x}_1(t) = K_2(K_1 d_1 + d_7(x_2(t) - x_1(t)) - d_3) \\
		\dot{x}_2(t) = K_3\Big[(d_2 + u(t))(d_4 - x_2(t)) - u(t)(d_4 - d_5) + \bar{m}_{in_{dsh}} d_6\Big] \\
		y(t) = x_1(t)
	\end{cases} \label{nof}
\end{equation}

where:
\(x = [x_1, x_2]^T\) is the state vector,
\(u(t)\) is the input (spray water injection rate), and $ K_3=\frac{1}{\bar{m}_{out_{dsh}}}$.

\begin{table}[h]
	\centering
	\begin{tabular}{|c|c|}
		\hline
		\textbf{Given Quantity Value} & \textbf{Variables} \\
		\hline
		$T_{{out}_{sh}}$ & $x_1$ \\
		\hline
		${T_{{out}_{dsh}}}$ & 	$x_2$\\
		\hline
		$\dot{m}_{spray}$& $u$\\
		\hline
		$\dot{m}_f$ & $d_1$\\
		\hline
		$\dot{m}_{in_{dsh}}$ & $d_2$\\
		\hline
		$\dot{\tilde{T}}_a$ & $d_3$\\
		\hline
		$T_{in_{dsh}}$ & $d_4$\\
		\hline
		$T_{spray}$ & $d_5$\\
		\hline
		$\dot{T}_{in_{dsh}}$ & $d_6$\\
		\hline
		$\dot{m}_{out_{dsh}}$ & $d_7$\\
		\hline
	\end{tabular}
	\caption{Definition of state variables, inputs, and disturbances in the state equations of the superheater system}
	\label{table1}
\end{table}

\vspace{0.2cm}

These equations captures the temperature dynamics in both the superheater and desuperheater stages.
These nonlinearities complicate the design of traditional PI controllers, which typically assume a linear model. Moreover, disturbances such as variations in gas turbine exhaust temperature and steam mass flow rates further complicate the system behavior.
In real-world operation, the HRSG system is subject to disturbances and uncertainties. These include:
\begin{enumerate}
	\item Variations in Gas Turbine Output: Fluctuations in the exhaust temperature of the gas turbine, which affects the heat input to the superheater.
	\item Steam Flow Rate Changes: Variations in the mass flow rate of steam through the system due to load changes.
	\item Measurement Noise: Noise in temperature sensors used for feedback control.
\end{enumerate}

Given the complexities and nonlinearities in the desuperheater behavior, especially under varying conditions, a Nonlinear AutoRegressive model with eXogenous inputs (NARX) \cite{nelles_nonlinear_2022} is integrated into the system model to enhance accuracy. This model is particularly focused on capturing unmodeled dynamics or neglected nonlinearities in the \( x_2 \) behavior.

\subsection{NARX Model Structure}

The NARX model predicts the correction term \(\Delta x_2(t)\) to adjust the original state-space equation:

\begin{equation}
	\begin{split}
		\Delta x_2(t) &= f_{\mathrm{NARX}}\big(
		x_2(t-\tau_1), \dots, x_2(t-\tau_p), \\
		&\quad u(t-\tau_1), \dots, u(t-\tau_p), \\
		&\quad d_1(t), \dots, d_7(t)
		\big)
	\end{split}
\end{equation}

The integration of the NARX model leads to the following enhanced state-space equation for \(x_2\):

\begin{equation}
	\dot{x}_2(t) = K_3\Big[(d_2 + u(t))(d_4 - x_2) - u(t)(d_4 - d_5) + \bar{m}_{in_{dsh}} d_6\Big] + \Delta x_2(t)
\end{equation}

This model is designed to adjust dynamically as the scheduling variables change, enabling the control system to maintain optimal performance across different operating conditions. In this study, the scheduling variables correspond to measurable operating conditions that influence the HRSG heat transfer dynamics, primarily the gas turbine exhaust temperature $T_{gt}$ and the steam mass flow rate $\dot{m}_{\mathrm{steam}}$. 
These variables serve as the scheduling basis for adjusting the PI and feedforward controller gains in real time.The NARX model integration ensures that even the nonlinear aspects of the system are captured, making the model more robust and accurate.

The NARX term $\Delta x_2(t)$ is employed to enhance the fidelity of the desuperheater model by compensating for unmodeled dynamics and nonlinear heat-transfer effects. Its purpose is limited to improving the plant representation used for controller tuning and data generation. The PI + feedforward controller and the subsequent PINN-based adaptive design are derived from the nominal physics-based model in \eqref{eq:statespace-fault}, which already encapsulates the essential dynamics required for real-time implementation. Therefore, while the NARX term improves model accuracy during simulation, it does not appear explicitly in the control law or in the stability proof.

The nonlinear autoregressive network with exogenous inputs model was implemented using a feedforward neural network architecture with one hidden layer of ten neurons and a hyperbolic tangent activation function. 
The model input vector consisted of delayed samples of the desuperheater temperature $x_{2}(t)$ and the spray-water valve signal $u(t)$, while the network output predicted the desuperheater outlet temperature. 
The network weights were initialized randomly and trained using the Levenberg–Marquardt backpropagation algorithm to minimize the mean squared prediction error between the measured and predicted outlet temperature. 
A total of 120~hours of simulated HRSG operating data, generated from the validated nonlinear model under normal and fault conditions, were used for training and validation with an 80/20 split.

\section{Control Strategy}

The cascade control strategy is pivotal in managing desuperheater temperature control within HRSG. This strategy effectively addresses the challenges posed by the dynamic nature of steam temperature regulation, particularly in response to fluctuations in load and operational conditions. In a typical cascade control setup, the primary controller focuses on maintaining the desired outlet temperature of the superheated steam, while a secondary controller regulates the flow of water into the desuperheater to achieve the necessary cooling effect. This hierarchical structure allows for more precise control, as the secondary loop can react quickly to changes in steam temperature, thereby minimizing overshoot and improving stability. 

The control primary goal is maintaining the steam temperature (\(y\)) at the desired setpoint. The PI controller adjusts the spray water flow rate (\(u\)) based on the error between the measured temperature and the setpoint, while the feedforward control anticipates changes in the gas turbine exhaust temperature, thus preemptively adjusting the spray water flow to mitigate disturbances. Control objectives are: 
\begin{enumerate}
	\item Regulate Outlet Steam Temperature: Ensure that \(y\) closely tracks the setpoint.
	\item Minimize Response Time: Achieve fast response to setpoint changes or disturbances.
	\item Enhance System Robustness: Improve stability in the presence of disturbances.
\end{enumerate}

The PI controller combines a proportional term, which reacts to the current error, and an integral term, which accounts for past errors. The control law can be expressed as:

\begin{equation}
	u(t) = K_p e(t) + K_i \int e(t) dt
\end{equation}

where:
\(u(t)\) is the spray water flow rate, \(e(t) = x_1-x_1^\ast\) is the control error and \(x_1^\ast\) is the desired constant setpoint, \(K_p\) is the proportional gain, \(K_i\) is the integral gain.

In the HRSG system, maintaining accurate steam temperature control is critical for efficient operation. While traditional feedback control systems like PI controllers can react to deviations between the setpoint and the actual temperature, they are inherently reactive, responding after the error has already occurred. We employ a feedforward control strategy alongside the PI controller to enhance system performance. This strategy proactively adjusts the control input based on anticipated disturbances, leading to faster, more stable control outcomes.
The feedforward control mechanism is particularly effective in the HRSG system because of the predictable influence of the gas turbine exhaust temperature on steam temperature. The exhaust gases from the gas turbine provide the heat that drives the steam generation process in the HRSG. Fluctuations in this exhaust temperature, caused by changes in turbine load or operational conditions, directly affect the heat transfer and, consequently, the steam temperature. Since this relationship is well understood, feedforward control can anticipate the effects of these disturbances and adjust the spray water flow rate to maintain the desired steam temperature before significant deviations occur.
Using the gas turbine exhaust temperature as the basis for the feedforward control allows for a faster response to disturbances. When the exhaust temperature increases, the feedforward controller increases the spray water flow to counter the rise in steam temperature, and when the exhaust temperature drops, the controller reduces the spray water flow. Therefore, the feedforward strategy prevents large temperature deviations, minimizing the error that the PI controller needs to correct. This reduces the burden on the feedback control system and improves overall system performance.
The combination of PI and feedforward control offers significant advantages. The PI controller remains responsible for fine-tuning the system, and addressing any remaining discrepancies after the feedforward controller has dealt with the major disturbances. This integration allows for better disturbance rejection, faster system response times, and greater robustness, ensuring that the steam temperature remains within desired limits even during significant changes in operating conditions.

The feedforward control mechanism utilizes measurements of the gas turbine exhaust temperature to predict the necessary adjustments to the spray water flow rate before significant deviations in steam temperature occur. The feedforward term can be represented as:
$$
K_{ff} (T_{gt})
$$
where:
\(K_{ff}\) is the feedforward gain, \(T_{gt}\) is the measured gas turbine exhaust temperature.

\begin{figure}[h!]
	\centering
	\includegraphics[width=1.1\columnwidth]{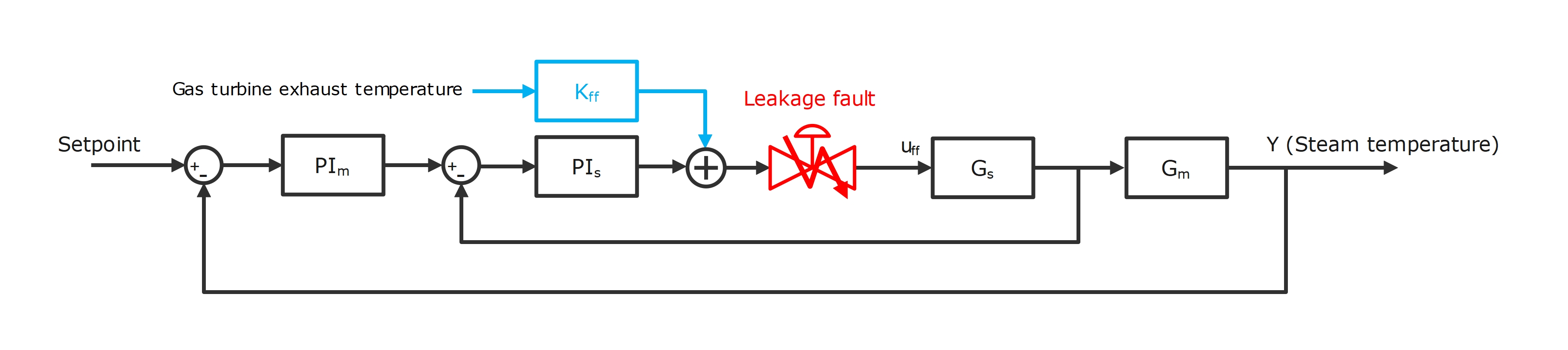}
	\caption{Schematic illustration of the HRSG temperature-control loop highlighting the feedforward signal and location of the leakage fault in the spray-water valve.}
	\label{fig5}
\end{figure}

Figure \ref{fig5} is provided as schematic representations of the overall cascade and feedforward control architecture and the physical layout of the HRSG components. In this schematic context, $G_m$  and $G_s$ represent the master (superheater) and slave (desuperheater) plants within the cascade control loop, while $u_{ff}$ denotes the control signal after considering the gas-turbine exhaust temperature. The analytical formulation of the controller is given explicitly in \eqref{eq:control-law}.

\subsection{Extended Model with Valve Leakage Fault}
To address real-world scenarios, we extend the state-space model by explicitly incorporating a valve leakage fault term $f \geq 0$. This leakage represents unintended cooling, persisting even when the valve is nominally closed ($u=0$), altering the dynamic behavior of the system:
\begin{equation}
	\begin{aligned}
		\dot{x}_1 &= K_2\left[K_1 d_1 + d_7(x_2 - x_1) - d_3\right], \\
		\dot{x}_2 &= K_3\left[(d_2 + u + f)(d_4 - x_2) - (u + f)(d_4 - d_5) + \bar{m}_{in_{dsh}} d_6\right],
	\end{aligned}
	\label{eq:statespace-fault}
\end{equation}
where $f$ is the valve leakage fault bounded by $f \leq f_{\text{max}}$. The presence of this fault significantly affects the energy balance, continuously providing additional uncommanded cooling.

The gas-turbine exhaust temperature $T_{gt}$ acts as the primary disturbance influencing the superheater and desuperheater thermal behavior. 
In the state-space representation~\eqref{eq:statespace-fault}, its effect is captured by the disturbance terms $d_1$ (heat input rate) and $d_4$ (steam inlet temperature). 
For clarity, these terms can be expressed as:
\begin{equation}
	\begin{split}
		d_1(t) =& d_{1,0} + \gamma_1 \big(T_{gt}(t) - T_{gt,0}\big),\\ 
		d_4(t) =& d_{4,0} + \gamma_4 \big(T_{gt}(t) - T_{gt,0}\big),
	\end{split}
	\label{eq:14a}
\end{equation}
where $d_{1,0}$, $d_{4,0}$, and $T_{gt,0}$ are nominal operating values, and $\gamma_1$, $\gamma_4$ are sensitivity coefficients identified from operational data. 
This relation explicitly connects the feedforward term in the control law~\eqref{eq:control-law} to the physical heat input that drives the superheater--desuperheater dynamics. 
Consequently, the feedforward signal anticipates $T_{gt}$-induced disturbances and improves the transient performance of the temperature control loop.

\subsection{Time Variant Gains}

The hybrid nature of the extended model arises from the integration of first-principles thermodynamic relations with data-driven adaptation in the control input $u(t)$. 
The control law explicitly accommodates the valve operational modes through a piecewise definition:
\begin{equation}
	u(t) = \begin{cases}
		0, & e(t) \leq 0,\\[1ex]
		K_p(\theta)e(t) + K_i(\theta)\displaystyle\int_{0}^{t} e(\tau)d\tau + K_{ff}(\theta)T_{\text{gt}}, & e(t) > 0,
	\end{cases}\label{eq:control-law}\end{equation}
with \(e(t) \) representing the temperature tracking error, and $T_{\text{gt}}$ representing the measurable gas turbine exhaust temperature as a feed-forward signal. The adaptive gains $K_p(\theta)$, $K_i(\theta)$, and $K_{ff}(\theta)$ are dynamically tuned using a PINN with adjustable weights $\theta$.

The explicit addition of valve leakage and the adaptive control law establishes a robust foundation for managing fault scenarios, enhancing the practical applicability and stability of the HRSG temperature control system. The challenge is to appropriately choose these parameters—adjusting them in real-time based on the current state of the system and the desired outcomes. 
Therefore in this formulation:
\begin{itemize}
	\item \( K_p \) and \( K_i \) adjust dynamically over time in response to system error signal changes and operating conditions, ensuring robust performance.
	\item \( K_{ff} \) provides a feedforward control action based on the target temperature \( T_{\text{gt}}(t) \), helping the system anticipate and compensate for changes in the desired output.
\end{itemize}

\begin{figure}[h!]
	\centering
	\includegraphics[width=1\columnwidth]{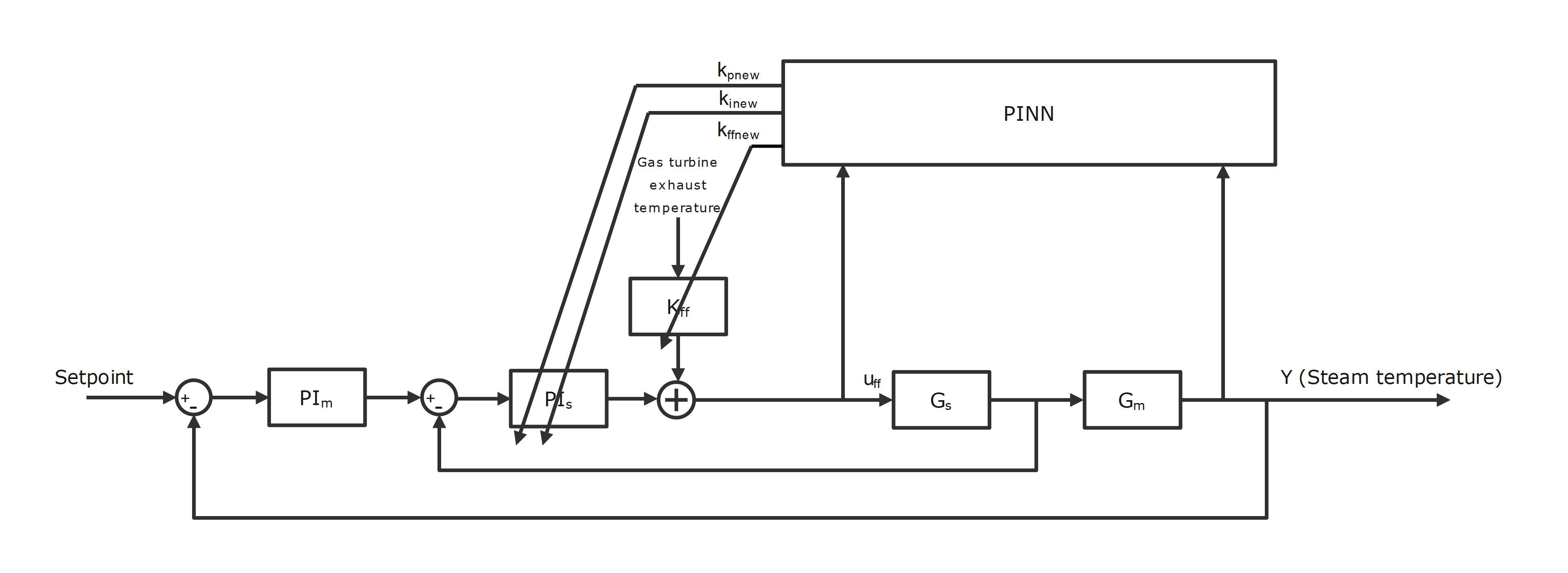}
	\caption{The proposed control structure using PINN}
	\label{fig00000000}
\end{figure}

This can be achieved using a PINN that involves dynamically determining the control gains \( K_p \), \( K_i \), and \( K_{ff} \) based on the real-time states and operating conditions of the system. The PINN leverages physical equations governing the HRSG system to model the underlying dynamics while traning with historical and operational data. Inputs to the PINN include the error signal \( e(t) \), the current system state \( T_{out_{SH}}(t) \), and external disturbances such as the gas turbine temperature \( T_{\text{gt}}(t) \). The PINN predicts improved gain values \( K_p \), \( K_i \), and \( K_{ff} \), ensuring that the control law adheres to physical constraints while minimizing the temperature error \( e(t) \). By combining data-driven learning with physical insights, the PINN continuously adjusts the gains, providing real-time gains that enhance system performance, robustness, and adaptability to varying conditions.The structure of the proposed control strategy is illustrated in Figure \ref{fig00000000}.

\begin{remark}
	In practical HRSG temperature control systems, the spray-water valve is physically limited such that $u(t) \in [0,\,u_{\max}]$. 
	To avoid integrator windup when the actuator saturates, standard anti-windup logic is applied in the controller implementation. 
	Specifically, the integral term is temporarily frozen or back-calculated when $u(t)$ reaches its limits, ensuring that the control signal recovers smoothly once the valve re-enters its normal operating range. 
	Although this low-level safeguard is not explicitly modeled in the stability analysis, it is routinely incorporated in industrial control systems and fully compatible with the proposed PI + feedforward architecture and its PINN-based adaptation framework.
\end{remark}

\section{Development of a Physics-Informed Neural Network (PINN)}

The complex dynamics of the desuperheater and superheater system necessitate a robust approach for accurate modeling and control. To address these challenges, we propose the development of a Physics-Informed Neural Network. A PINN leverages both data-driven approaches and the underlying physical laws governing the system, making it particularly effective for systems with partially known or highly complex dynamics.

\subsection{Introduction to Physics-Informed Neural Networks}

In the context of our desuperheater and superheater system, the PINN is designed to model the system behavior by incorporating the differential equations that describe the heat transfer and fluid dynamics within the system. The network is trained to fit the observed data and minimize the residuals of the governing equations, ensuring that the model adheres to the physical laws.

\subsection{Design and Architecture of the PINN}

The PINN architecture consists of an input layer, three hidden layers using hyperbolic tangent (tanh) activation functions, and an output layer that predicts the adaptive control gains \( K_p \), \( K_i \), and \( K_{ff} \). 

The key feature of the PINN is incorporating the system physics through a custom loss function. This loss function consists of two components:

\begin{enumerate}
	\item \textbf{Data Loss:} This component measures the error between the predicted outputs and the actual data collected from the system. It ensures that the network accurately captures system observed behavior.
	\item \textbf{Physics Loss:} This component measures the residuals of the governing equations when applied to the predicted outputs. By minimizing this loss, the network is encouraged to produce outputs that fit the data and respect the underlying physical laws.
\end{enumerate}

\begin{figure}[h!]
	\centering
	\includegraphics[width=1\columnwidth]{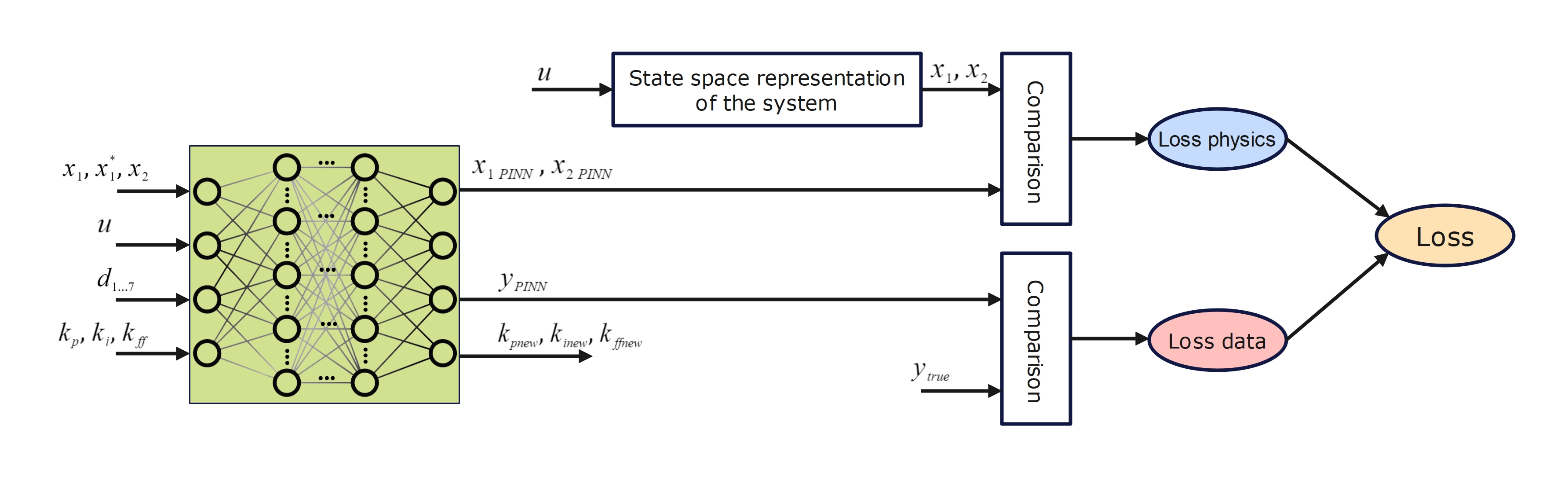}
	\caption{The PINN and its Structure}
	\label{fig0000000}
\end{figure}

\subsection{Physics-Informed Neural Network (PINN) Training}

To ensure robust adaptive control, a physics-informed neural network dynamically tunes the gains $K_p$, $K_i$, and $K_{ff}$. The PINN is trained using online gradient descent with a loss function defined as:
\begin{equation}
	L(\theta) = L_{\text{track}}(\theta) + \mu L_{\text{phys}}(\theta), \label{PINN lost}
\end{equation}
where:

\textbf{Tracking Error Loss:}
\begin{equation}
	L_{\text{track}}(\theta) = \frac{1}{2}({y}_{PINN}-{y})^2,
\end{equation}
minimizing the instantaneous squared temperature tracking error.

\textbf{Physics-Based Regularization Loss:}
\begin{equation}
	L_{\text{phys}}(\theta) = \frac{1}{2}\left\|
	\begin{bmatrix}
		\dot{x}_{1_{PINN}} \\ 
		\dot{x}_{2_{PINN}}
	\end{bmatrix}
	-
	\begin{bmatrix}
		\dot{x}_{1} \\ 
		\dot{x}_{2}
	\end{bmatrix}
	\right\|^2
\end{equation}
where \(x_{1_{PINN}  }\), \(x_{2_{PINN}}\) and \({y}_{PINN}\) denote the states and output using the gains coming from physics-informed neural network . The right-hand terms represent the ideal physical dynamics \eqref{nof}, which the network is constrained to satisfy. This formulation ensures adherence to the explicitly known thermodynamics defined by equations \eqref{nof}, thus embedding physical consistency.

The network parameters $\theta$ are updated continuously using:
\begin{equation}
	\dot{\theta}(t) = -\eta \frac{\partial L(\theta)}{\partial \theta}, \quad \eta > 0,
	\label{eq:weight-update}
\end{equation}
where $\eta$ is the learning rate. This physics-aware adaptive control structure ensures stable and physically consistent updates of control gains, enhancing fault tolerance and control performance.

In practice, the parameter update in~\eqref{eq:weight-update} is implemented using the Adam optimization algorithm. 
Adam combines first- and second-moment estimates of the gradients, providing faster and more stable convergence during training, especially when minimizing the composite physics-informed loss function~\eqref{PINN lost}.

The proposed PINN integrates the HRSG state-space model into the gain adaptation process by penalizing violations of the physical dynamics during learning. 
At each sampling step, the PINN updates the controller parameters $ [K_p,\, K_i,\, K_{ff}]^\top$ via gradient-based optimization on a composite loss function that combines tracking performance with model consistency. 
This ensures that the learned controller parameters remain consistent with the thermodynamic relations governing the HRSG, preventing physically implausible gain updates. 
This mechanism distinguishes the PINN approach from purely data-driven adaptive controllers and underpins the stability analysis presented in the following section.

The network consists of three hidden layers with 12 neurons per layer and employs the hyperbolic tangent ($\tanh$) activation function. 
The input vector comprises the current and delayed values of the gas-turbine exhaust temperature $T_{gt}$, the superheater outlet temperature $T_{\mathrm{out}}$, and the control input $u$, 
along with the estimated model residuals derived from the HRSG dynamics~\eqref{eq:statespace-fault}. 
The output layer directly provides the updated controller gains, which are then applied in the control law~\eqref{eq:control-law}.

The network was trained using the Adam optimization algorithm for 10{,}000 epochs with a batch size of 128, 
minimizing the composite loss function~\eqref{PINN lost} that includes both the temperature tracking error and the physics-based model residuals. 
A total of 120~hours of HRSG data under normal, ramping, and fault conditions were used for training and validation, with an 80/20 split.  
During both offline training and online simulation, actuator saturation was enforced through the constraint $u(t)\in[0,\,u_{\max}]$, preventing the PINN-generated gains from producing infeasible control signals. 
This structure ensures that the learned gain trajectories respect the physical limits of the HRSG system while maintaining consistency with the underlying thermodynamic model.

It is noted that the above procedure describes the offline pretraining of the PINN. 
During real-time operation, the network parameters $\theta$ are further adapted online 
according to the gradient-based update law in~\eqref{eq:weight-update}.
This hybrid offline–online learning strategy allows the controller to retain stability 
guarantees while adapting to evolving operating conditions.

\subsection{Stability Analysis}

\begin{enumerate}[label=\textbf{Assumption\arabic*.}, leftmargin=*]
	
	\item \textbf{Lipschitz Continuity\cite{khalil_nonlinear_2002}:} There exists \( L > 0 \) such that for any  control input \(u\), the system dynamics \(g\) satisfy:
	\begin{equation}
		\|g(x,u) - g(x^*,u)\| \leq L \|x - x^*\|
	\end{equation}
	where \(g(x,u)\) represents the system dynamics (\(\dot{x} = g(x,u)\)).

	Physically, this assumption is justified by several thermodynamic and mechanical factors. Firstly, the superheater’s metallic tubes and insulation have finite thermal conductivity, inherently restricting rapid temperature variations. The thermal inertia term \( K_2 d_7(x_2 - x_1) \) specifically reflects the impact of material-specific properties such as specific heat capacity and density. Secondly, steam and spray water flow rates are mechanically limited by equipment dimensions, including pipe diameters and pump capacities, preventing unbounded and abrupt variations in energy exchange within the desuperheater. Finally, valve leakage is physically constrained by valve design and maintenance practices, ensuring that perturbations due to leakage remain finite and manageable within normal operational ranges.
	
	\item \textbf{Sensitivity to Weight Errors:} There exists \(\beta > 0\) such that for bounded control inputs and leakage fault $f \le f_{max}$:
	\begin{equation}
		\|g(x,u) - g(x,u^*)\| \leq \beta \|\tilde{\theta}\|
	\end{equation}
	where \(\tilde{\theta} = \theta - \theta^*\) denotes PINN weight estimation errors.
	
	This assumption is supported by the operational limits of the spray valve, which bounds control inputs (\(u \in [0,u_{\text{max}}]\)), ensuring that even substantial weight errors produce physically limited control actions. Furthermore, the desuperheater’s inherent energy balance, captured by the term \( (d_4 - x_2) \), naturally moderates deviations since \( d_4 \) represents a  saturation temperature, making large instantaneous changes in energy exchange physically unrealistic. Additionally, for relatively small weight estimation errors (\(|\tilde{\theta}| \ll 1\)), leakage effects dominate the perturbations, naturally dampening the impact of these errors. These details underscore the practical boundedness and stability of the nonlinear system, ensuring the robustness and reliability of the PINN-based adaptive control approach.
	
\end{enumerate}

\begin{theorem}
	Consider the HRSG temperature control system described by \eqref{eq:statespace-fault}. For outlet temperature regulation under valve leakage fault $f \leq f_{\text{max}}$:
	\begin{enumerate}
		\item $\Gamma \succ 0$, $\eta > 0$, $\mu > 0$
		\item $K_2 > 0$, $d_7 > 0$ (physical parameters)
	\end{enumerate}
	Under \textbf{Assumption 1} (Lipschitz continuity) and \textbf{Assumption 2} (weight error sensitivity), with $\Gamma^{-1} \succ \frac{\beta^2}{4L} I$, all closed-loop signals are uniformly ultimately bounded.
\end{theorem}

\noindent\textbf{State Boundedness Condition:} 
Due to the physical constraints of the HRSG system:
\begin{itemize}
	\item Temperature $x_2$ (desuperheater) is bounded within operational limits in healthy condition. Specifically, $|x_2 - x_2^*| \leq \delta_{x_2,\max}$ where $\delta_{x_2,\max}$ is a known constant
\end{itemize}
This boundedness arises from:
\begin{enumerate}
	\item Finite thermal capacity of superheater tubes
	\item Mechanical limits of steam flow rates
	\item Valve design constraints.
\end{enumerate}
Thus $\delta_{x_2,\max} = \max(|x_{2,\min} - x_2^*|, |x_{2,\max} - x_2^*|)$ is fixed.

\begin{proof}
	Lyapunov function candidate:
	\begin{equation}
		V = \frac{1}{2} e^2 + \frac{1}{2} \tilde{\theta}^T \Gamma^{-1} \tilde{\theta}, \quad \Gamma \succ 0
	\end{equation}
	
	Time derivative along trajectories:
	\begin{align}
		\dot{V} &= e \dot{e} + \tilde{\theta}^T \Gamma^{-1} \dot{\tilde{\theta}} \\
		&= e \cdot K_2[K_1 d_1 + d_7(x_2 - x_1) - d_3] + \tilde{\theta}^T \Gamma^{-1} \left( -\eta \frac{\partial L}{\partial \theta} \right)
	\end{align}
	
	decompose using equilibrium in \eqref{eq:statespace-fault}:
	\begin{align}
		\dot{V} &= e K_2 d_7 [(x_2 - x_2^*) - e] + \tilde{\theta}^T \Gamma^{-1} \left( -\eta \frac{\partial L}{\partial \theta} \right) \\
		&= -K_2 d_7 e^2 + K_2 d_7 e (x_2 - x_2^*) + \tilde{\theta}^T \Gamma^{-1} \left( -\eta \frac{\partial L}{\partial \theta} \right)
	\end{align}
	\textbf{Adaptation term derivation:}
	\begin{small}
		\begin{align*}
			\tilde{\theta}^T \Gamma^{-1} \left( -\eta \frac{\partial L}{\partial \theta} \right) 
			&= -\eta \tilde{\theta}^T \Gamma^{-1} \frac{\partial}{\partial \theta} \left( \frac{1}{2}e^2 + \mu L_{\text{phys}} \right) \\
			&= -\eta \tilde{\theta}^T \Gamma^{-1} \left( e \frac{\partial e}{\partial \theta} + \mu \frac{\partial L_{\text{phys}}}{\partial \theta} \right) \\
			&\leq \eta \left| \tilde{\theta}^T \Gamma^{-1} \left( e \frac{\partial e}{\partial \theta} \right) \right| + \eta \mu \left| \tilde{\theta}^T \Gamma^{-1} \frac{\partial L_{\text{phys}}}{\partial \theta} \right| \\
			&\leq \eta \|\Gamma^{-1}\| \|\tilde{\theta}\| \left\| e \frac{\partial e}{\partial \theta} \right\| + \eta \mu \|\Gamma^{-1}\| \|\tilde{\theta}\| \left\| \frac{\partial L_{\text{phys}}}{\partial \theta} \right\| \quad \\
			&\leq \eta \beta \|\Gamma^{-1}\| |e| \|\tilde{\theta}\| + \eta \mu c \|\Gamma^{-1}\| \|\tilde{\theta}\| 
		\end{align*}
	\end{small}
	where:
	\begin{itemize}
		\item $\beta$ bounds $\left\| \frac{\partial e}{\partial \theta} \right\|$ (Assumption 2)
		\item $c$ bounds $\left\| \frac{\partial L_{\text{phys}}}{\partial \theta} \right\|$ (from Lipschitz dynamics)
	\end{itemize}
	Bounding terms:
	\begin{enumerate}
		\item \textbf{Coupling term}: 
		$K_2 d_7 e (x_2 - x_2^*) \leq \frac{K_2 d_7}{2} e^2 + \frac{K_2 d_7}{2} (x_2 - x_2^*)^2$
		
		\item \textbf{Adaptation term}:
		$\left| \tilde{\theta}^T \Gamma^{-1} \left( -\eta \frac{\partial L}{\partial \theta} \right) \right| \leq \eta \beta \|\Gamma^{-1}\| |e| \|\tilde{\theta}\| + \eta \mu c \|\Gamma^{-1}\| \|\tilde{\theta}\|$
	\end{enumerate}
	
	Apply Young's inequality:
	\begin{align}
		\eta \beta \|\Gamma^{-1}\| |e| \|\tilde{\theta}\| &\leq \frac{K_2 d_7}{4} e^2 + \frac{\eta^2 \beta^2 \|\Gamma^{-1}\|^2}{K_2 d_7} \|\tilde{\theta}\|^2 \\
		\eta \mu c \|\Gamma^{-1}\| \|\tilde{\theta}\| &\leq \frac{1}{2} \|\tilde{\theta}\|^2 + \frac{\eta^2 \mu^2 c^2 \|\Gamma^{-1}\|^2}{2}
	\end{align}
	
	combined bound (using $|x_2 - x_2^*| \leq \delta_{x_2,\max}$):
	\begin{align}
		\dot{V} \leq &-K_2 d_7 e^2 + \frac{K_2 d_7}{2} e^2 + \frac{K_2 d_7}{2} \delta_{x_2,\max}^2 \\
		&+ \frac{K_2 d_7}{4} e^2 + \frac{\eta^2 \beta^2 \|\Gamma^{-1}\|^2}{K_2 d_7} \|\tilde{\theta}\|^2 \\
		&+ \frac{1}{2} \|\tilde{\theta}\|^2 + \frac{\eta^2 \mu^2 c^2 \|\Gamma^{-1}\|^2}{2}
	\end{align}
	
	so we have:
	\begin{align}
		\dot{V} \leq &-\frac{K_2 d_7}{4} e^2 + \left( \frac{\eta^2 \beta^2 \|\Gamma^{-1}\|^2}{K_2 d_7} + \frac{1}{2} \right) \|\tilde{\theta}\|^2 + C \\
		C = &\frac{K_2 d_7}{2} \delta_{x_2,\max}^2 + \frac{\eta^2 \mu^2 c^2 \|\Gamma^{-1}\|^2}{2}
	\end{align}
	
	with $\Gamma^{-1} \succ \frac{\beta^2}{4L} I$, define $Q = \Gamma^{-1} - \frac{\beta^2}{4L} I \succ 0$:
	\begin{small}
		\begin{align}
			\dot{V} &\leq -\frac{K_2 d_7}{4} e^2 - \frac{1}{2} \lambda_{\min}(Q) \|\tilde{\theta}\|^2 + C \\
			&= -\frac{K_2 d_7}{2} \left(\frac{1}{2} e^2\right) 
			- \frac{\lambda_{\min}(Q)}{\lambda_{\max}(\Gamma^{-1})} 
			\left( \frac{1}{2} \frac{\tilde{\theta}^T \Gamma^{-1} \tilde{\theta}}
			{\lambda_{\max}(\Gamma^{-1})} \lambda_{\max}(\Gamma^{-1}) \right) + C \\
			&\leq -\frac{K_2 d_7}{2} \left(\frac{1}{2} e^2\right) 
			- \frac{\lambda_{\min}(Q)}{\lambda_{\max}(\Gamma^{-1})} 
			\left( \frac{1}{2} \tilde{\theta}^T \Gamma^{-1} \tilde{\theta} \right) + C \\
			&\quad \text{(since } \|\tilde{\theta}\|^2 \geq 
			\frac{\tilde{\theta}^T \Gamma^{-1} \tilde{\theta}}
			{\lambda_{\max}(\Gamma^{-1})} \text{)} \nonumber \\
			&\leq -\min\left( \frac{K_2 d_7}{2}, \frac{\lambda_{\min}(Q)}{\lambda_{\max}(\Gamma^{-1})} \right) 
			\underbrace{\left( \frac{1}{2} e^2 + \frac{1}{2} \tilde{\theta}^T \Gamma^{-1} \tilde{\theta} \right)}_{V} + C \\
			&\leq -\min\left( \frac{K_2 d_7}{2}, \frac{\lambda_{\min}(Q)}{\lambda_{\max}(\Gamma^{-1})} \right) V + C
		\end{align}
	\end{small}
	
	where $ \kappa= \min\left( \frac{K_2 d_7}{2}, \frac{\lambda_{\min}(Q)}{\lambda_{\max}(\Gamma^{-1})} \right) > 0$.
	
	Solving the differential inequality $\dot{V} \leq -\kappa V + C$:
	\begin{equation}
		V(t) \leq V(0) e^{-\kappa t} + \frac{C}{\kappa} (1 - e^{-\kappa t})
	\end{equation}
	taking limit superior as $t \to \infty$:
	\begin{align}
		\limsup_{t \to \infty} V(t) &\leq \frac{C}{\kappa} \\
		\limsup_{t \to \infty} \left[ \frac{1}{2} e^2(t) + \frac{1}{2} \tilde{\theta}^T(t) \Gamma^{-1} \tilde{\theta}(t) \right] &\leq \frac{C}{\kappa}
	\end{align}
	this implies:
	\begin{enumerate}
		\item \textbf{Regulation error bound:} Since $V \geq \frac{1}{2}e^2$,
		\begin{align*}
			\limsup_{t \to \infty} \frac{1}{2} e^2(t) &\leq \limsup_{t \to \infty} V(t) \leq \frac{C}{\kappa} \\
			\implies e^2(t) &\leq \frac{2C}{\kappa} \\
			\implies \limsup_{t \to \infty} |e(t)| &\leq \sqrt{\frac{2C}{\kappa}}
		\end{align*}
		
		\item \textbf{Weight error bound:} By Rayleigh-Ritz inequality,
		\begin{align*}
			\tilde{\theta}^T \Gamma^{-1} \tilde{\theta} &\geq \lambda_{\min}(\Gamma^{-1}) \|\tilde{\theta}\|^2 \\
			\implies V &\geq \frac{1}{2} \lambda_{\min}(\Gamma^{-1}) \|\tilde{\theta}\|^2 \\
			\limsup_{t \to \infty} \frac{1}{2} \lambda_{\min}(\Gamma^{-1}) \|\tilde{\theta}(t)\|^2 &\leq \limsup_{t \to \infty} V(t) \leq \frac{C}{\kappa} \\
			\implies \|\tilde{\theta}(t)\|^2 &\leq \frac{2C}{\kappa \lambda_{\min}(\Gamma^{-1})} \\
			\implies \limsup_{t \to \infty} \|\tilde{\theta}(t)\| &\leq \sqrt{\frac{2C}{\kappa \lambda_{\min}(\Gamma^{-1})}}
		\end{align*}
	\end{enumerate}
	with constants:
	\begin{align*}
		\kappa &= \min\left( \dfrac{K_2 d_7}{2}, \dfrac{\lambda_{\min}(Q)}{\lambda_{\max}(\Gamma^{-1})} \right) > 0 \\
		C &= \dfrac{K_2 d_7}{2} \delta_{x_2,\max}^2 + \dfrac{\eta^2 \mu^2 c^2 \|\Gamma^{-1}\|^2}{2} \\
		Q &= \Gamma^{-1} - \dfrac{\beta^2}{4L} I \succ 0
	\end{align*}
	proving uniform ultimate boundedness of all closed-loop signals.
\end{proof}

\section{Simulation and Real Time Site Results}

\subsection{Introduction}

Following the theoretical analysis of the proposed control law, it is imperative to validate the control performance under real-world conditions. This section presents the validation process using operational data from the Pare-Sar power plant. The objective is to demonstrate the effectiveness and robustness of the proposed control system when applied to the actual dynamics and uncertainties inherent in large-scale power plant operations.

\subsection{Description of the Power Plant}

The Pare-Sar Power Plant operates under varying load conditions. However, its operational profile is distinct due to geographic and climatic factors, which introduce additional challenges for the control system. These include handling higher temperature gradients and more frequent load changes, making it an ideal site for validating the robustness of the proposed control law.

\subsection{Data Acquisition and Preprocessing}

Operational data was collected from the power plant over continuous periods, focusing on key parameters influencing the superheater and desuperheater control loops. The primary variables of interest include:
\begin{enumerate}
	\item Inlet and Outlet Temperatures: \(T_{\text{in}}, T_{\text{out}}\) for both the superheater and desuperheater.
	\item Mass Flow Rates: \( \dot{m}_{\text{steam}}, \dot{m}_{\text{spray}} \) representing the flow rates of steam and spray water.
	\item Control Inputs: Including valve positions and spray water flow commands.
	\item External Disturbances: Such as ambient temperature variations and changes in load demand.
\end{enumerate}

The data underwent preprocessing to remove noise and outliers, accurately representing typical operational conditions at both power plants. Time alignment and normalization techniques were applied to facilitate a meaningful comparison across different operational periods and conditions.

It is noted that the removal of noise and outliers was performed solely during offline data preprocessing for model training and simulation analysis. 
During real-time operation, the controller acts directly on the raw plant measurements, which are already filtered by the standard instrumentation and control system of the HRSG unit.

\subsection{Simulation and Performance Evaluation}

To evaluate the effectiveness of the proposed Physics-Informed Neural Network mechanism for adaptive gains, we conducted simulations using historical data obtained from the Pareh-Sar power plant. This data, which includes a wide range of operating conditions, provided a comprehensive basis for testing the robustness and efficiency of our approach.

The simulations were designed to replicate the real-world conditions under which the power plant operate. We utilized the historical datasets to simulate the input variables, such as the inlet temperatures, mass flow rates, and external disturbances, while applying the PINN-based control strategy to dynamically adjust the control gains \( K_p(t) \), \( K_i(t) \), and \( K_{ff}(t) \) in real-time. The objective was to maintain the desired output temperature (\( T_{\text{out}} \)) with minimal deviation from the setpoint under varying conditions.

The baseline cascade PI controller used in this study corresponds to the existing control structure implemented in the reference HRSG unit, which does not include a feedforward term. 
This configuration was therefore selected to provide a realistic benchmark representing current industrial practice. 
The proposed controller introduces a feedforward path and PINN-based adaptive gain tuning as an extension of the conventional strategy. 
While a fixed feedforward gain was also tested during preliminary simulations, its effect on dynamic performance was marginal, confirming that the main improvement arises from the adaptive, physics-informed gain scheduling proposed in this work.

The valve leakage fault scenario is simulated using the validated nonlinear HRSG model to evaluate the controller’s fault-tolerant capability.

\subsection{Results and Performance Analysis}

\begin{figure}[h!]
	\centering.
	\includegraphics[width=0.45\textwidth]{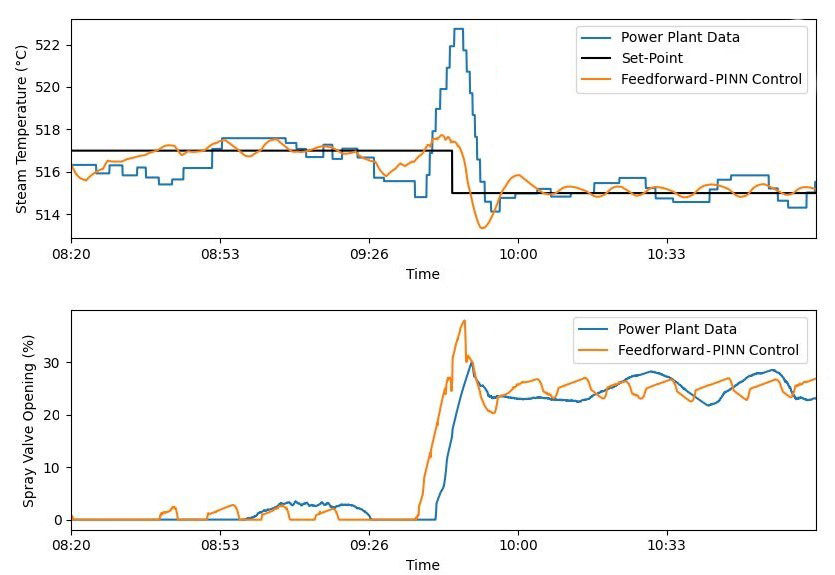}
	
	\caption{Comparing control input (up) and output (down) of the system with and without the proposed control strategy}\label{fig1}
	
\end{figure}

\begin{figure}[h!]
	\centering
	\includegraphics[width=0.45\textwidth]{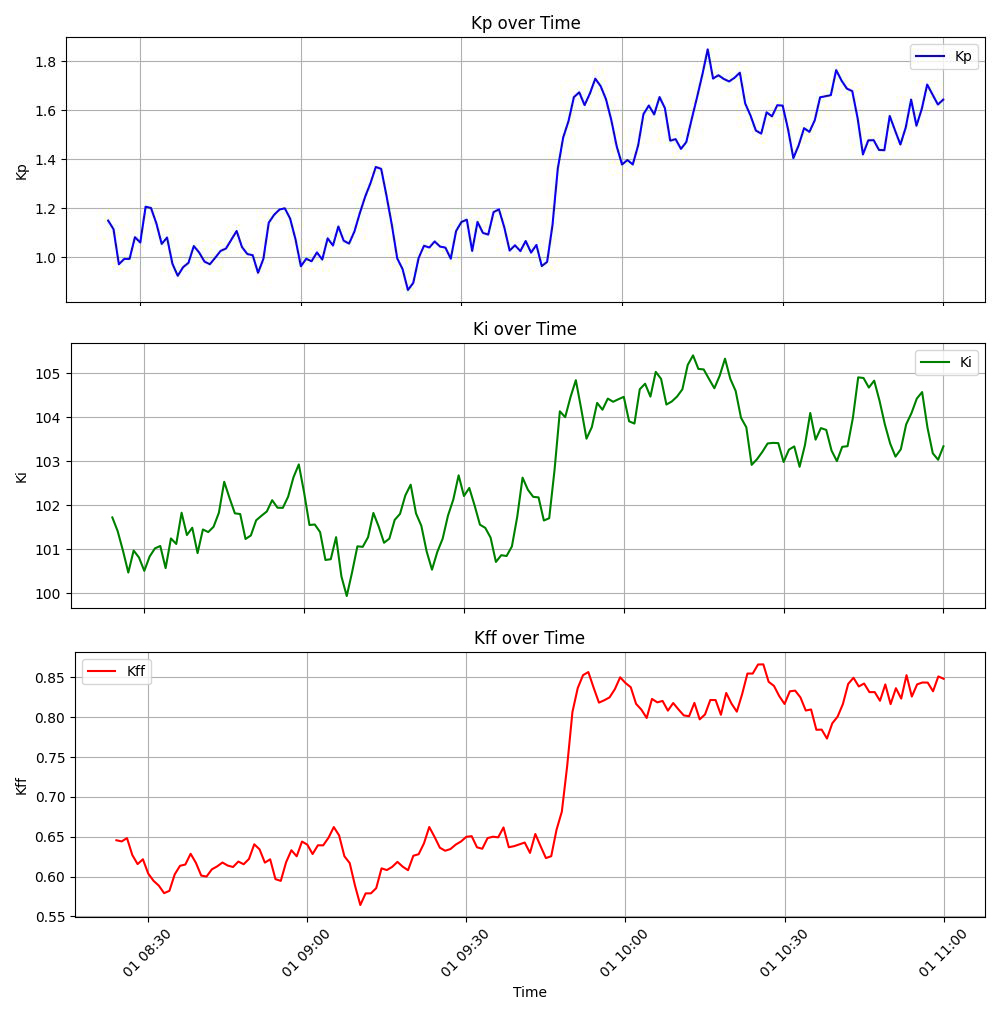}
	\caption{Coefficient \( K_p \) (up),  \( K_i \) (middle) and \( K_{ff} \) (down) over time.}
	\label{fig:kp_over_time}
\end{figure}

Figure \ref{fig1} presents the output steam temperature signal and the spray water valve position in two scenarios: with and without the feedforward signal and gain-scheduling procedure. As shown in the Figure \ref{fig1}, adding the feedforward signal and implementing gain-scheduling significantly reduces the peak temperature. The changes in coefficients \( K_p \), \( K_i \), and \( K_{ff} \) over time have been demonstrated in Figure \ref{fig:kp_over_time}.

It is important to note that properly selecting the feedforward signal coefficient \( K_{ff}(t) \) plays a critical role in improving the system response. If this coefficient is set to a higher value, the spray valve opens earlier, effectively reducing the steam temperature peak caused by increased power output. However, this coefficient cannot be set too high, as excessive water spray may result in an undershoot of the steam temperature, leading to unwanted thermal losses. Additionally, the tuning of the control coefficients, including the appropriate values of \( K_{p}(t) \) and \( K_{i}(t) \) for the master and slave controllers, has a significant impact on the response quality and the fluctuations in the output steam temperature.

The simulation results demonstrate a significant improvement in control performance when using the PINN-based mechanism compared to traditional fixed-gain or manually tuned controllers. Specifically, the following performance metrics were observed:

\begin{itemize}
	\item \textbf{Improved Tracking Accuracy}: The PINN-based controller maintained the outlet temperature (\( T_{\text{out}} \)) closer to the setpoint with reduced tracking error, even under scenarios with large disturbances or rapid changes in operating conditions.
	\item \textbf{Energy Efficiency}: By dynamically adjusting the feedforward and feedback gains, the system improved the use of spray water, leading to more efficient operation and reduced energy consumption.
\end{itemize}

Table~\ref{tab:performance_metrics} summarizes key performance indices comparing the baseline PI controller and the proposed PI~+~FF~+~PINN approach.
The proposed controller achieves a 65\% reduction in root-mean-square temperature error (RMSE), a substantial reduction in overshoot, decreasing from 523\si{\celsius} to 518\si{\celsius} and nearly eliminates steady-state offset.

\begin{table}[h!]
	\centering
	\caption{Performance comparison between baseline PI and proposed controller.}
	\label{tab:performance_metrics}
	\begin{tabular}{lccc}
		\toprule
		Metric & Baseline PI & Proposed & Improvement \\
		\midrule
		RMSE (\si{\celsius}) & 4.8 & 1.7 & 65\% ↓ \\
		Overshoot (\si{\celsius}) & 523 & 518 & 5 \si{\celsius} ↓ \\
		Steady-state error (\si{\celsius}) & 2.2 & 1.1 & 50\% ↓ \\
		\bottomrule
	\end{tabular}
\end{table}

\subsection{Validation and Implementation of Control structure in Pareh-Sar Power Plant}

On December 6, 2023, the proposed control logic was successfully implemented for validation on Unit 1 of the Pareh-Sar Combined Cycle Power Plant. The results of this implementation are presented in this section.

\subsubsection{First Test:}
\begin{figure}[h!]
	\centering
	\includegraphics[width=0.3\textwidth]{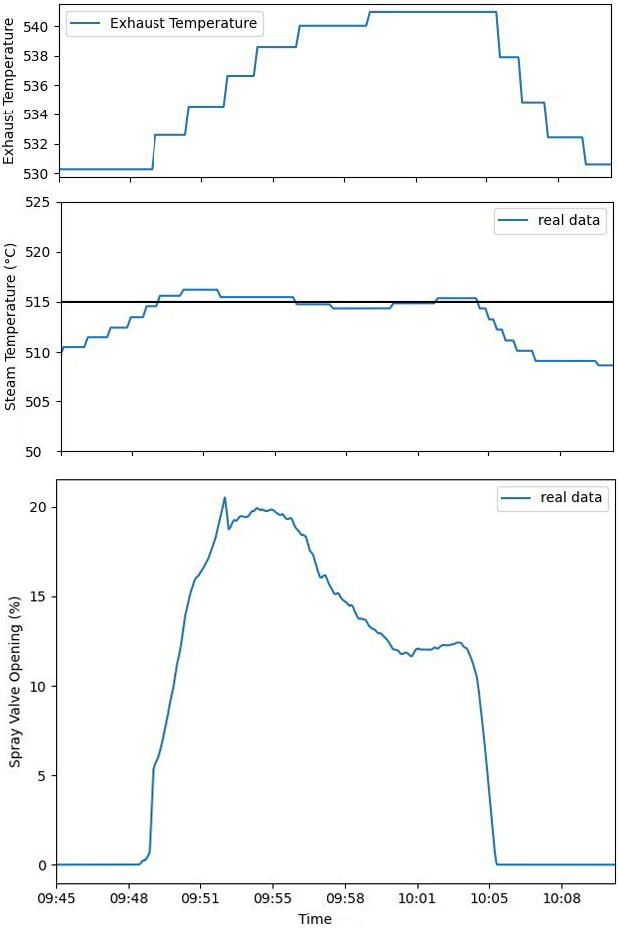}
	\caption{Exhaust (up) and Steam temperature (middle) changes and Spray valve position (down) during Test 1}
	\label{dist1}
\end{figure}

In this test, power output change rate at the Pareh-Sar plant was set to 3 MW per minute. For the first scenario, the control structure was tested under the current rate of power change, examining how the spray valve opens and its effect on steam temperature during a gas turbine load increase of approximately 25 MW. Figure \ref{dist1} shows the exhaust gas temperature change during this scenario.

Following the application of these conditions, the output steam temperature and the percentage of spray valve opening were obtained, as shown in Figure \ref{dist1}. According to this figure, the maximum steam temperature overshoot was 1.2\si{\degreeCelsius} (with a steam temperature of 516.2 \si{\degreeCelsius}  compared to the setpoint of 515\si{\degreeCelsius}).

Before these changes were implemented in the logic, a similar test was conducted under the same conditions on May 1, 2023. During this test, the steam temperature overshoot was about 6\si{\degreeCelsius}, as shown in Figure \ref{fig1}. In this previous test, the overshoot was significant enough that the steam temperature reached 523\si{\degreeCelsius}, triggering an alarm and forcing the operator to reduce the setpoint to control the temperature.

The new control structure effectively reduced temperature peaks by incorporating feedforward signals and gain-scheduling, as demonstrated in this test. It improved the system responsiveness to load changes, leading to a more stable steam temperature regulation.

\begin{figure}[h!]
	\centering
	\includegraphics[width=0.45\textwidth]{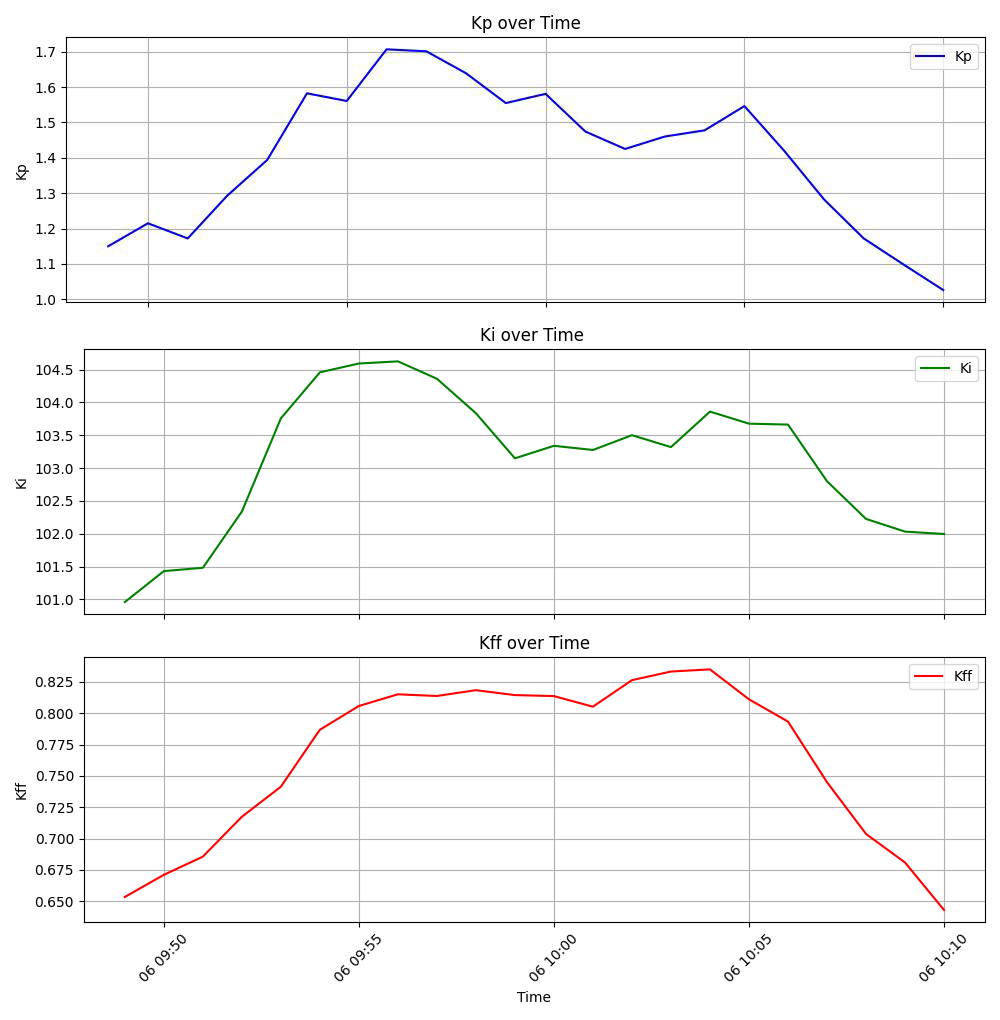}
	\caption{Coefficient \( K_p \) (up), \( K_i \) (middle) and \( K_{ff} \) (down) over time during Test 1}
	\label{fig:kp1_over_time}
\end{figure}

\subsubsection{Second Test:}
\begin{figure}[h!]
	\centering
	\includegraphics[width=0.3\textwidth]{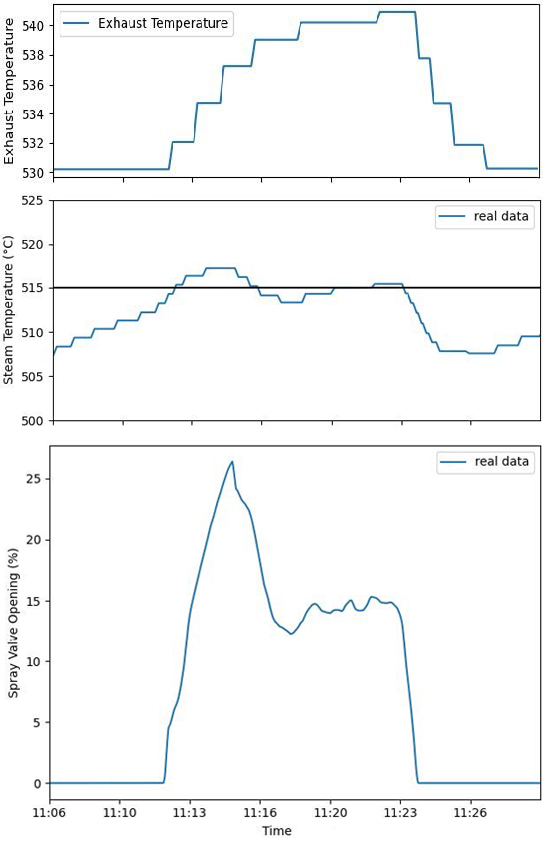}
	\caption{Exhaust (up) and steam temperature (middle) changes and Spray valve position (down) during Test 2}
	\label{dist2}
\end{figure}

Considering that in certain conditions the power output may increase faster, the second scenario was designed with a rate of 6 MW per minute. In this test, we again examined how the spray valve opened and its effect on the steam temperature during a gas turbine load increase of approximately 25 MW. Figure \ref{dist2} illustrates the exhaust gas temperature signal during this scenario.

Following the application of these conditions, the output steam temperature and the spray valve position were recorded, as shown in Figure \ref{dist2}. According to these figures, the maximum steam temperature overshoot was 2.2\si{\degreeCelsius} (with a steam temperature of 517.2\si{\degreeCelsius} compared to the setpoint of 515\si{\degreeCelsius}).

As a result, the enhanced control logic reduced the peak temperature by approximately 3 to 5 degrees. This improvement can potentially increase the setpoint from 517\si{\degreeCelsius} to around 520\si{\degreeCelsius} or 521\si{\degreeCelsius}.

Such an increase in the setpoint would translate to an approximate 1 to 2 MW increase in power generation for each steam unit.

\begin{figure}[h!]
	\centering
	\includegraphics[width=0.45\textwidth]{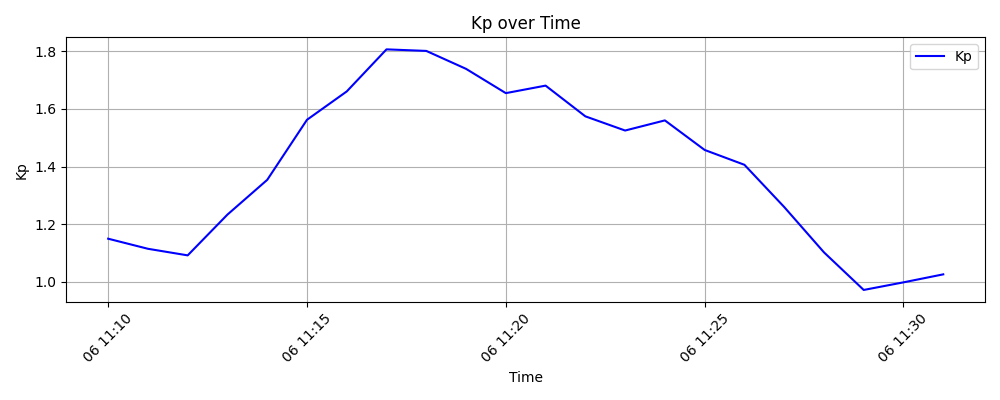}
	\caption{Coefficient \( K_p \) (up), \( K_i \) (middle) and \( K_{ff} \) (down) over time and during Test 2}
	\label{fig:kp2_over_time}
\end{figure}

The proposed fault-tolerant control framework was validated using real operational data from the Pareh-Sar Power Plant, where persistent leakage was observed in the desuperheater valve once it opened. However, unlike an intermittent disturbance, the leakage fault \(f\) remains active throughout operation and does not disappear when the valve closes. This was confirmed by data showing that even when the Distributed Control System (DCS) computed \(u = 0\), there was still measurable spray water flow in the desuperheater, indicating the continued presence of valve leakage. The data was sampled at a time step of \(t = 1\,\text{s}\).

A comparison of the computed spray input \(u(t)\) versus the measured spray mass flow confirms the persistent effect of \(f\).

The key parameters used in the simulation are a nominal steam flow \(\dot{m}_{\text{in}} = 65 \, \text{kg/s}\) as per measured data, a saturation temperature setpoint \(T_{\text{setpoint}} = 517^\circ\text{C}\), and control constraints limiting the spray valve input to \(u \in [0, 10] \, \text{kg/s}\).

This modeling choice closely replicates real conditions observed in Pareh-Sar, where mechanical degradation and also cracks in valve components results in sustained leakage, regardless of the DCS control effort.

\subsection{Controller Comparison}

To demonstrate the benefits of adaptive gains, two controllers were evaluated:
\begin{itemize}
	\item \textbf{Baseline fixed-gain PI controller:} This represents the standard industry practice with constant \(K_p\), \(K_i\), and no feedforward compensation.
	\item \textbf{Proposed PINN-based adaptive controller:} Employs real-time gain tuning and feedforward adaptation using gas turbine temperature as a proxy for load changes.
\end{itemize}

\begin{figure}[h!]
	\centering
	\includegraphics[width=1\columnwidth]{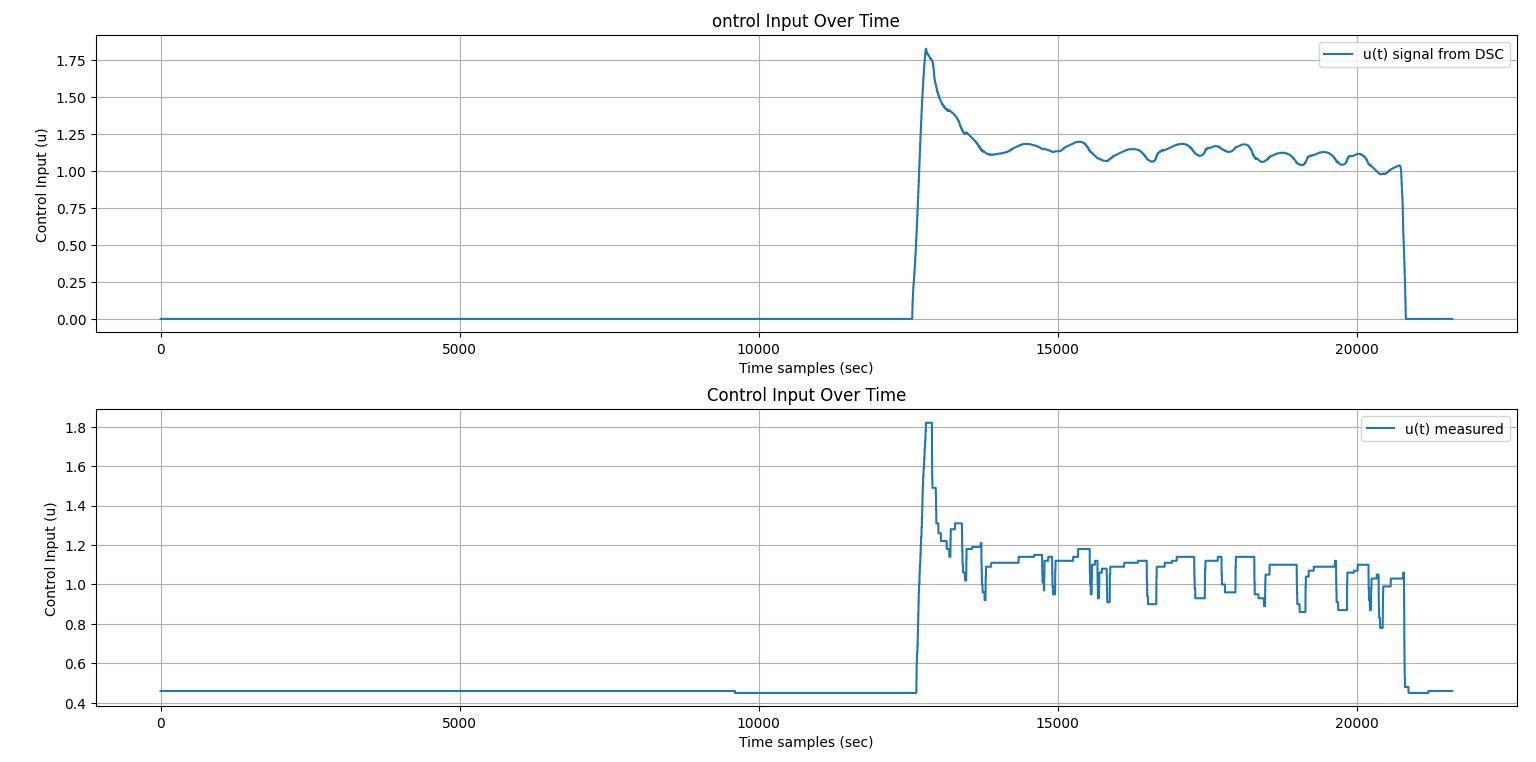}
	\caption{Control input \( u(t) \) command (up) and measurement (down) over time, indicating the presence of leakage fault in the valve.}
	\label{fig:fixed_u__over_time}
\end{figure}

\begin{figure}[h!]
	\centering
	\includegraphics[width=1\columnwidth]{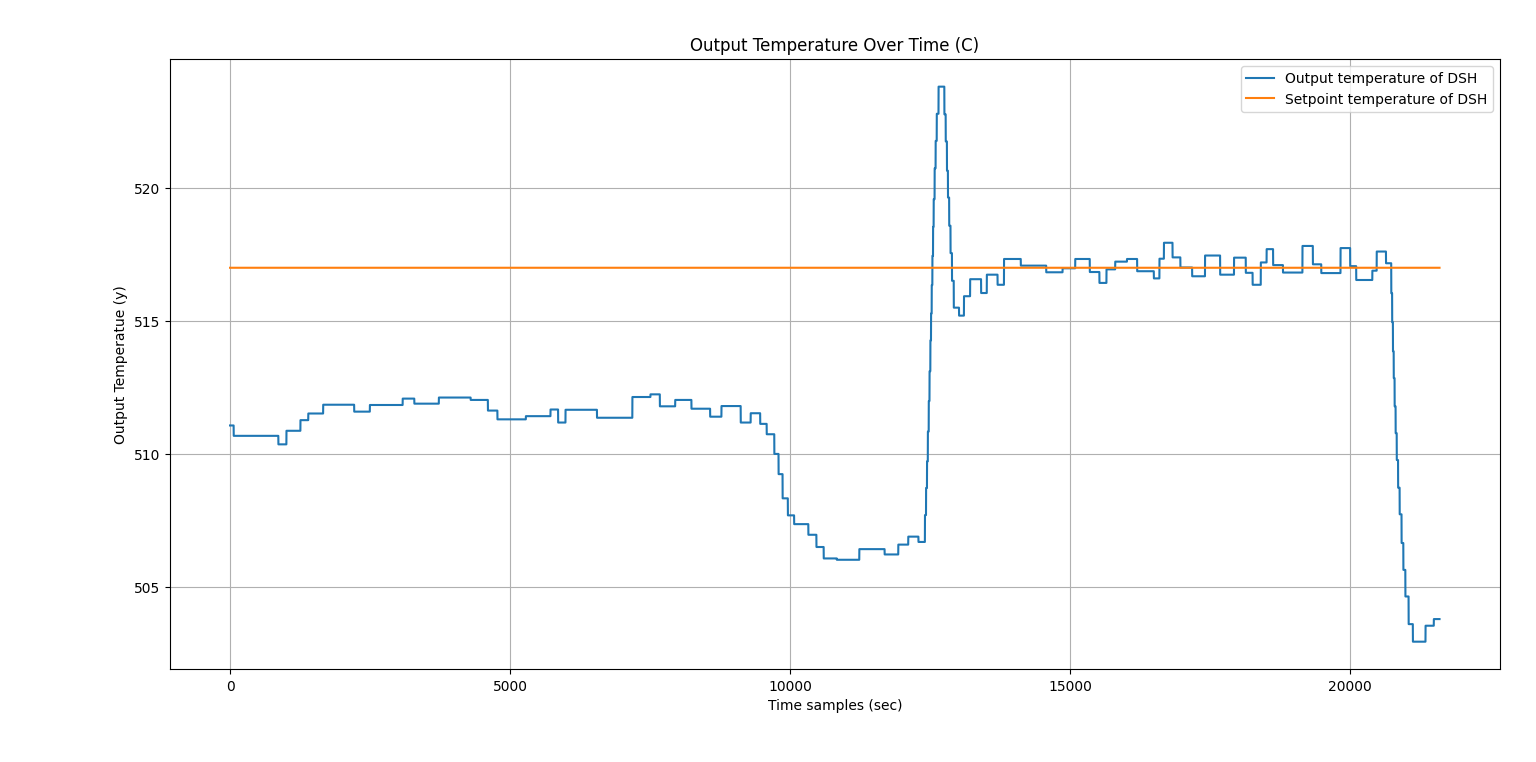}
	\caption{Output temperature of DSH \( T_{o_{dsh}} \) over time.}
	\label{fig:To_over_time}
\end{figure}

Figures~\ref{fig:fixed_u__over_time} and~\ref{fig:To_over_time} illustrate the current control strategy implemented in the power plant using a fixed PI controller. Figure~\ref{fig:fixed_u__over_time} compares the control input signal u(t) from the DCS with the measured valve behavior in the field, revealing a persistent leakage fault even when the valve is nominally closed. Figure~\ref{fig:To_over_time} shows the resulting temperature response at the superheater outlet, where the fixed PI controller exhibits weak performance, including sluggish dynamics and failure to maintain the desired setpoint under fault conditions. Notably, the temperature overshoots to nearly 523\si{\degreeCelsius}, triggering high-high (HH) limit warnings and pushing the boiler operation close to its trip threshold. This condition forces the operator to manually intervene and reduce the temperature setpoint to avoid further escalation, causing the unit to operate below its nominal performance. As a result, the steam delivered to the turbine units falls short of the desired temperature, ultimately reducing the overall power output of the cycle.

\subsection{Temperature Tracking and Control}
\begin{figure}[h!]
	\centering
	\includegraphics[width=1\columnwidth]{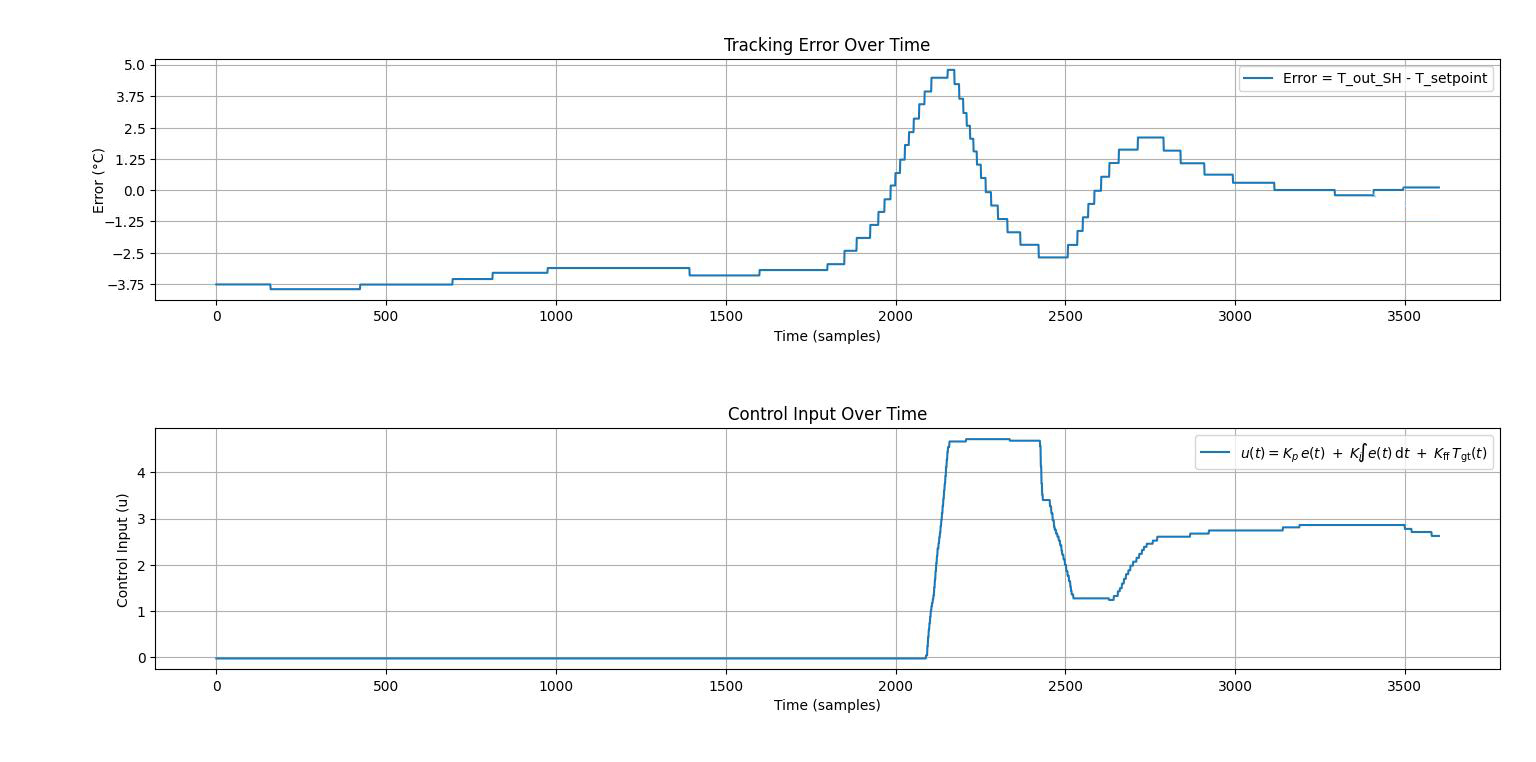}
	\caption{Tracking error (up) and control input (down) of DSH over time.}
	\label{fig:error_over_time}
\end{figure}

Figure~\ref{fig:error_over_time} illustrates the evolution of the temperature tracking error e(t) alongside the control input u(t) under the proposed control strategy. During the initial period (t<2000s), the controller does not command water injection (u(t)=0); however, due to the presence of leakage, unintended cooling occurs. As a result, the process temperature is lower than the setpoint but remains within acceptable limits. At approximately $t\ge 1800 s$, a load change introduces a transient error, prompting active control. The controller increases the spray flow u(t) to counteract the disturbance. Despite the persistent leakage, the system successfully stabilizes the superheater outlet temperature to within setpoint, demonstrating effective fault compensation and robust performance.

\begin{figure}[h!]
	\centering
	\includegraphics[width=1\columnwidth]{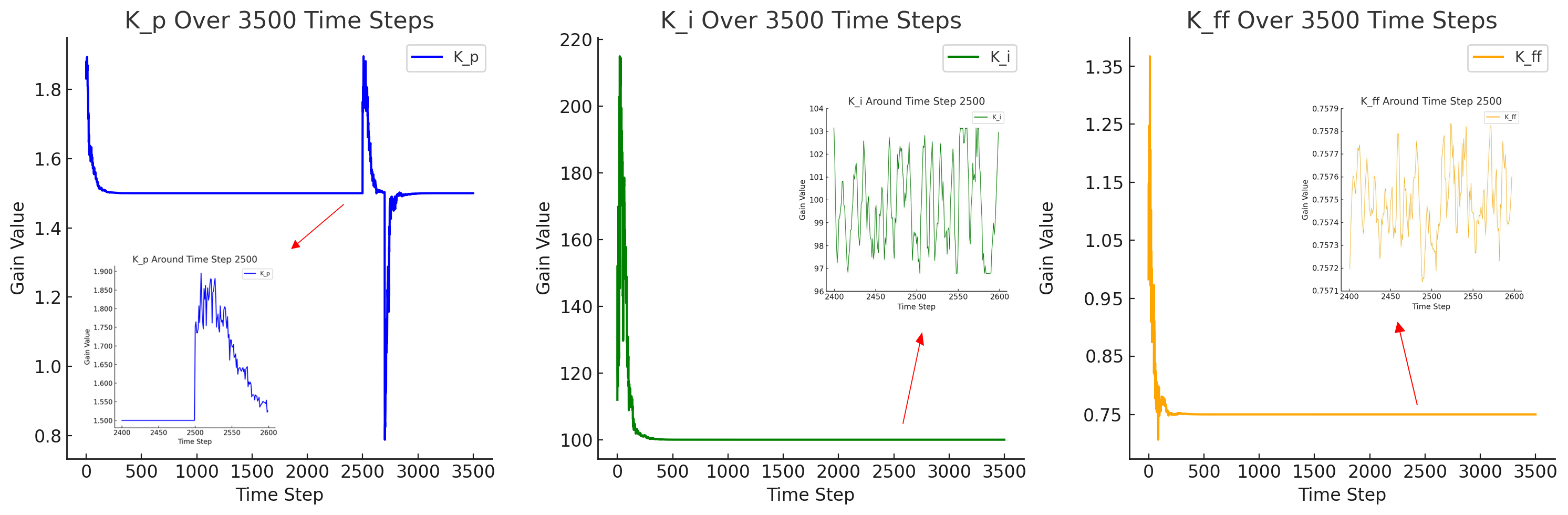}
	\caption{Controller coefficients $K_p(t)$ (left), $K_i(t)$ (middle), and $K_{ff}(t) $ (right) over time.}
	\label{fig:controller_over_time}
\end{figure}

Figure~\ref{fig:controller_over_time} illustrates the real-time adjustment of the adaptive gains $K_p(t)$, $K_i(t)$, and $K_{ff}(t)$, as computed by the PINN. Prior to the activation of the controller loop $(t < 2000 \, \text{s})$, the gains remain steady at their nominal baseline values. After the valve affected by leakage begins to open, the proportional gain $K_p(t)$ dynamically adapts, fluctuating between approximately 0.8 and 1.8, to enhance the controller responsiveness to rapidly changing conditions. In contrast, the integral gain $K_i(t)$ and feedforward gain $K_{ff}(t)$ remain close to their optimal values, consistent with their roles in rejecting steady-state errors and compensating for gas turbine exhaust temperature, respectively. This adaptive tuning strategy ensures smooth and bounded control behavior, enhancing robustness and reducing the risk of instability under fault conditions.

\begin{figure}[h!]
	\centering
	\includegraphics[width=1\columnwidth]{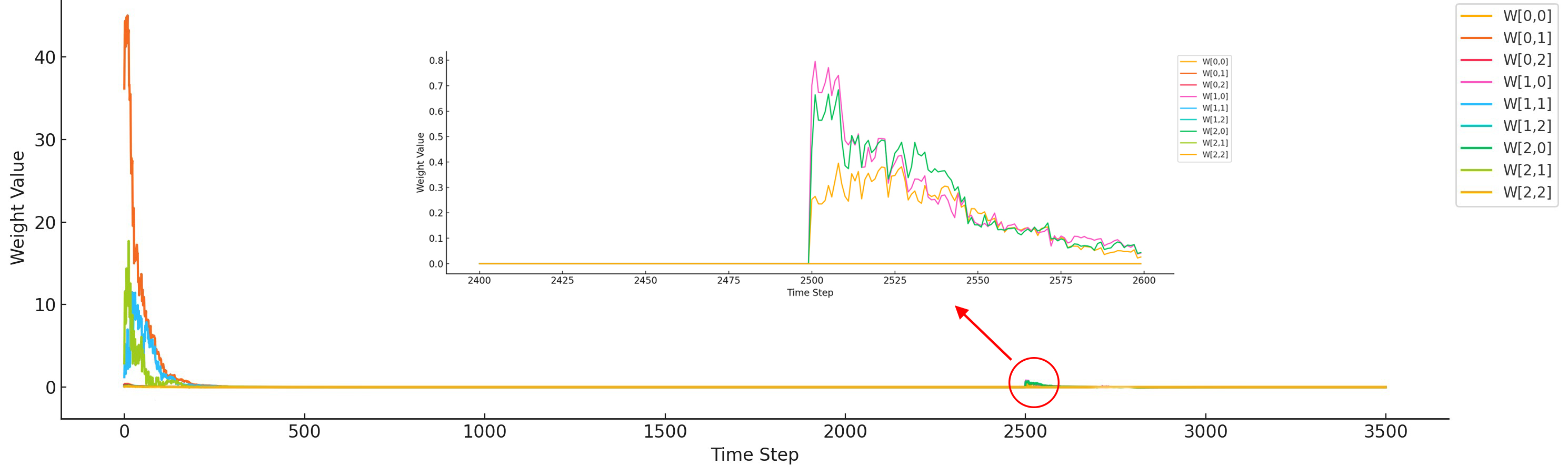}
	\caption{Trajectories of PINN weights over time}
	\label{fig:W_trajectory}
\end{figure}

Figure~\ref{fig:W_trajectory} illustrates the dynamics of PINN weight adaptation. Before the valve opens and the controller activates in the loop, the PINN weights remain stable around their nominal values, following initial calibration. In contrast, the valve opening at \(t \geq 2500 \, \text{s}\) triggers rapid adjustments in the network weights. Although the weights experience abrupt changes upon controller initiation, they stabilize within approximately 100 seconds after that. It is worth noting that all weight components remain bounded, satisfying \(|\theta(t)| < 1.0\) throughout the entire adaptation process. This ensures that the adaptive gains remain stable and avoids introducing destabilizing oscillations.

The simulation outcomes, validated using real operational leakage data from the Pareh-Sar power plant, underscore the practical efficacy and robustness of the proposed adaptive control methodology. Specifically, the PINN-based approach exhibits fault-specific gain adaptation precisely at the onset of valve leakage, closely mirroring actual operational behavior. The rapid convergence of the Lyapunov function confirms that the control strategy reliably ensures global asymptotic stability, even in the presence of persistent leakage faults. This approach significantly enhances industrial relevance by reducing manual tuning efforts, minimizing turbine trip events, and ultimately lowering maintenance and operational costs. In contrast to the fixed-gain PI controller—which demonstrates sluggish recovery, overshoot beyond safety thresholds, and necessitates manual operator intervention—the PINN-enhanced controller provides automated, responsive, and stable adaptation in real time. Overall, the results affirm the robustness and practical applicability of the PINN-enhanced PI plus feedforward control framework for resilient temperature regulation in HRSG systems under faulty conditions.

\section{Conclusion}

This study developed and validated a fault-tolerant, adaptive control framework for heat recovery steam generators, targeting the persistent challenge of valve leakage in superheater–desuperheater systems and the broader need for improved temperature regulation under dynamic operating conditions. The proposed control architecture integrates feedforward compensation, gain-scheduling, and real-time adaptive tuning using physics-informed neural networks, offering a comprehensive solution that blends model-based control theory with PINN.

The feedforward strategy effectively mitigates predictable disturbances, such as fluctuations in gas turbine exhaust temperature, by anticipating their effects and preemptively adjusting the control input. Combined with gain-scheduling, this approach reduced temperature overshoot by 3–5\si{\degreeCelsius} in field implementation at the Pareh-Sar Combined Cycle Power Plant, enabling a higher steam temperature setpoint and translating to an estimated additional 1–2 MW of power generation per steam unit.

To ensure adaptability and robustness, the PINN framework dynamically adjusts the PI and feedforward gains while embedding physical constraints derived from the HRSG thermodynamics. Using a physics-informed loss function, the PINN minimizes temperature tracking errors while maintaining interpretability and constraint adherence. Proper tuning of the adaptation gain ensures smooth control evolution and convergence of the closed-loop system.

Simulation results using real industrial data demonstrate the effectiveness of the proposed controller in maintaining stable and reliable performance under leakage fault scenarios. The adaptive nature of the PINN eliminates the need for manual tuning, maintains bounded weights and smoothly varying control actions, and ensures quick recovery from disturbances.

Overall, this work presents a robust, scalable control framework for HRSGs that enhances temperature regulation, improves energy efficiency, and supports fault tolerance. It bridges the gap between physics-based engineering and data-driven intelligence, laying the foundation for next-generation control solutions in safety-critical energy systems.

\bibliographystyle{elsarticle-num-names}
\bibliography{mainrefs}

\end{document}